\documentstyle[epsf,epsfig,here,rotating]{article}
\newlength{\reducedwidth}
\title{Physics Potential at FNAL
with Stronger Proton Sources}
\author{
\\\\
G. Barenboim$^a$,
A.~de Gouv\^ea$^a$,
T. Dombeck$^a$,\\
N. Grossman$^a$, 
D. Harris$^a$, 
D. Michael$^b$,
M. Szleper$^c$,\\
M. Velasco$^c$,
S. Werkema$^a$ \\\\
\small{$^a$Fermilab,}
\small{$^b$California Institute of Technology,}
\small{$^c$Northwestern University}
}

\def\ie{{\sl i.e.}}

\def\numu{\nu_{\mu}}

\def\nue{\nu_e}

\def\nutau{\nu_{\tau}}

\def\mutoe{\numu \rightarrow \nue}

\def\mutotau{\numu \rightarrow \nutau}

\def\gtwid{\mathrel{\raise.3ex\hbox{$>$\kern-.75em\lower1ex\hbox{$\sim$}}}}
\def\ltwid{\mathrel{\raise.3ex\hbox{$<$\kern-.75em\lower1ex\hbox{$\sim$}}}}

\def\beq{\begin{equation}}
\def\eeq{\end{equation}}
\def\bea{\begin{eqnarray}}
\def\eea{\end{eqnarray}}
\def\bq{\begin{quote}}
\def\eq{\end{quote}}
\def\ben{\begin{enumerate}}
\def\een{\end{enumerate}}

\def\ie{{\it i.e.}}

\def\lesssim{\mathrel{\mathpalette\vereq<}}
\def\gtrsim{\mathrel{\mathpalette\vereq>}}
\makeatletter
\def\vereq#1#2{
\lower3pt\vbox{\baselineskip1.5pt \lineskip1.5pt
\ialign{$\m@th#1\hfill##\hfil$\crcr#2\crcr\sim\crcr}}}
\makeatother

\newcommand{\text}[1]{\mbox{#1}}

\begin{document}

\begin{titlepage}
\begin{center}
\hfill    FERMILAB-FN-720\\
\hfill nuhep-exp/2002-03\\
\end{center}

\vskip 2cm

\begin{center}
{\large \bf Physics Potential at FNAL
with Stronger Proton Sources}
\end{center}
\vskip 1cm
\begin{center}
G. Barenboim$^a$,
A.~de Gouv\^ea$^a$,
T. Dombeck$^a$,\\
N. Grossman$^a$,
D. Harris$^a$,
D. Michael$^b$,
M. Szleper$^c$,\\
M. Velasco$^c$,
S. Werkema$^a$ \\
\vskip 1.0cm
\small{$^a$\em Fermi National Accelerator Laboratory \\
 P.O. Box 500, Batavia, IL 60510, USA}
\vskip 0.5cm
\small{$^b$ \em California Institute of Technology High Energy Physics, \\
Charles C.  Lauritsen Laboratory, Pasadena,  CA 91125, USA }
\vskip 0.5cm
\small{$^c$ \em  Northwestern University  \\
Department of Physics \& Astronomy \\
2145 Sheridan Rd, Evanston, IL 60208, USA}
\end{center}

\vspace{3.5cm}
\begin{abstract}

This document is the second in a series of reports on the exciting
physics that would be accessible at Fermilab in the event of an
upgraded proton source.  Where the first report covered a broad
range of topics, this report focuses specifically on three areas
of study:  there are brief discussions on the new measurements one could
make in both the neutron and anti-proton sectors, and then a detailed
discussion of what could be achieved in the neutrino
oscillation sector using an upgraded proton source to supply the
NuMI beamline with more protons.  
If one places a new detector
optimized for $\nu_e$ appearance at a new location slightly off the
axis defined by 
the MINOS experiment, that new experiment would be
ideal for making the next important steps in lepton flavor studies,
namely, the search for $\nu_\mu \to \nu_e$ at the atmospheric
mass splitting, and CP violations.
The report concludes with a summary of proton economics and demands
for increased proton intensity for the Booster and Main Injector: 
what the proton
source at Fermilab can currently supply, and what adiabatic
changes could be implemented to boost the proton supply on the way
from here to a proton driver upgrade.

\end{abstract}
\vskip 2cm
\end{titlepage}

\newpage
\tableofcontents

\vfill\eject

\section{Goals and Charge}
\par
In January of 2001 a study on the physics potential of a new high
intensity proton source facility ($\sim$0.5MW)\cite{pdriver} at Fermilab
was requested by Mike Shaevitz. The idea was that such a
facility could be the basis of a high quality U.S. based physics
program, to be operated  between  the second half of this decade and the
commissioning of the next major U.S. facility.  By that summer,
 the  working group had concluded that a rich and coherent program in
``Flavor Physics" could be achieved\cite{booster}, and that
the Proton Driver\,(PD) would have a large impact on a variety
of programs that are currently being pursued at Fermilab using the existing
machines as well as the potential for new facilities.
Experiments that would clearly benefit from a new PD
included  among others, the neutrino experiments covering
both the hot topic of oscillations (e.g. Minos and BooNE) as well as
non-oscillation physics. The possible new facilities are a neutron
spallation source and  a high intensity low energy muon source.
These facilities would be
competitive  and complementary to other existing or proposed facilities,
and would attract new users to Fermilab.

\par
In December 2001 by Mike~Shaevitz
and Steve~Holmes requested to re-invigorate the PD
studies.  Steve Holmes was interested in pursuing an
updated machine study with the goal of having a new design in time
for the ICFA Workshop on High Intensity High Brightness
Hadron Beams that took place at Fermilab on April~8-12 of~2002\cite{icfa}.
Mike~Shaevitz
requested a strong and detailed updated physics study to be presented
first at the same ICFA workshop and have final report  by  the PAC
scheduled for June,~2002.

\par
The new goals were to identify a range of accelerator configurations that
could provide 0.5-2\,MW of average beam power at 8-120~GeV, describe the
physics program that could be supported and identify required beamlines
and detectors.
The scope of this study was to be limited to three strong
physics topics that could be pursued with a new PD.  Conceptual
designs for the beam lines, targets and detectors were required,
as well as consideration of a physics program based on both a stand alone
source and an upgraded Main Injector.

\par
This document is the response to the request from Shaevitz and Holmes
from the physics working group\footnote{The findings from the machine
working group can be found in~\cite{Fermilab-TM-2169}.}. Even though we
focused on the physics program, we also include a discussion of proton
economics at Fermilab in order to shows that
the physics goals considered will require a significant increase
in the proton resources at Fermilab.  Only with an upgraded proton source,
such as the one described in~\cite{Fermilab-TM-2169} would we be able
to achieve the physics objectives listed here, and
fully capitalize on the Fermilab program already in place.

\section{Summary}
\par
There are many benefits that would come from a Proton Driver upgrade.
In this document we focus on three physics areas: neutrino
oscillations, static neutron properties and CPT tests with
anti-protons.  In all three areas Fermilab can make unique
and important physics contributions.

However, there is no doubt that neutrino physics is going through a
revolutionary moment, and that  neutrino masses and mixings
are already giving us new insights into the origin of flavor. In 
addition, given  the evidences for neutrino mass, leptogenesis 
is gaining momentum as the origin of cosmic baryon asymmetry, and therefore
CP violation in the lepton sector must be tested.
For these reasons, and many others, we have made 
our main emphasis neutrino oscillation physics, which
is the first topic covered. In the near future, at Fermilab we 
will definitively open~(or close) the door to new physics~\cite{nos} 
in the neutrino sector by confirming~(or not) the LSND anomaly~\cite{LSND} 
with MiniBooNE data~\cite{miniboone}. In case that LSND is confirmed we need 
even more protons than what we have available in the current Booster.
The physics case that should follow, in the case that MiniBoone
confirms LSND  and in  the availability of a new PD was already discussed
in detailed in our previous report~\cite{booster}. 
Here we focus on the experiments that are critical should LSND not be
confirmed.

More detailed studies of the
other two topics, static neutron properties and CPT tests with
anti-protons,  should come at a later date.

{\em Neutrino Oscillations:}
We describe in detail a proposal for an {\em off-axis} neutrino
experiment utilizing the NuMI beamline.  As we will show in the
following sections, this and the next generation of neutrino
experiments can measure
most of the leptonic mixing angles and mass-squared differences
from which it will be possible to infer the neutrino mass hierarchy.
The detector design will ensure the observation of 
$\nu_\mu \rightarrow \nu_e$ 
appearance, if the last unmeasured angle in the leptonic 
mixing matrix~($|U_{e3}|^2, \theta_{13}$) is large enough. 
 Depending on the
outcome, it will be possible to observe CP violations, if they
exist, by taking data with an anti-neutrino beam.  
At that stage of the program, the key to success 
is the higher proton flux afforded by the Proton Driver.
We show that this project can be staged to match Fermilab resources.

{\em Neutron Static Properties:}
The electric dipole moment~(EDM) of the neutron could provide
a window on new physics, as has been appreciated for
many years.  The current measurements constrain important
parameters in Supersymmetry, for example.  The next generation
of experiments will likely require a pulsed source of neutrons
in order to control systematics. 
As observed already in earlier reports on the proton 
driver~\cite{booster,pdriver}, a world class spallation
neutron source could be built at Fermilab utilizing the
Proton Driver which would have unique capabilities and
compliment other facilities elsewhere.

{\em CPT Tests with Anti-protons:}
General interest in tests of CPT symmetry is growing as
experimental efforts improve and theoretical possibilities
for CPT violation are recognized.  Some of the most important
work has been done with anti-protons at CERN, using the
AD (`Anti-proton Decelerator'). 
This program will be terminated, however, as CERN focuses
its resources on the LHC.  At FNAL we already have an anti-proton
source, therefore Fermilab is good candidate for the next generation
of anti-proton and anti-Hydrogen experiments.  The program can start
with the current facilities, and ultimately be enhanced with
the Proton Driver.

\subsection{Neutrino Program}

{\it An off-axis experiment in combination with a PD upgrade  would  be capable of observing both
$\mutoe$ and $\mutotau$ appearance, and $\nu_\mu$ disappearance.
In addition,  the experiment  will be able to 
measure  precisely the atmospheric parameters, measure or make
stringent exclusions of $|U_{e3}|^2~(\theta_{13})$,
and ultimately search for CP violation in the lepton sector and determine
the mass hierarchy once the new Proton Driver becomes available. 
Several detector technologies with the required physics specifications  
are available at varying cost and level of complexity.
Modest changes are required for the NuMI beam line, which is
already approved and currently under construction.}\\

We describe  a search for neutrino oscillations  in 
the atmospheric neutrino allowed region. The OFF-MINOS (Off-axis 
detector -- Main Injector Neutrino Oscillation Search)
program could  have two phases. The first phase  assumes 
a total of   125 kTon/years\footnote{Time scales are defined in term of 
the Snowmass-year ($1\times 10^7$ s). We need   five years 
to complete the program with a detector with a  20 kTon fiducial mass with 
NuMI ``as is", or a 5 kTon fiducial mass detector with a PD upgrade. The same 
program  can be completed  in just 1.5 years in case that we have   
both a PD upgrade and a 20 kTon  detector.}, and we will be able to:
 
\begin{itemize}
\item Extend the search for $\mutoe$ appearance for a factor bigger than 20  
 past the current experimental bound  at a  three sigma level.

\item Obtain an exclusion for $\mutoe$ appearance for $|U_{e3}|^2>0.0015$,  
 at a  two sigma level. This is approximately twenty times beyond 
current limits and almost an order of magnitude better than the MINOS
expectation if no signal is observed.
 
\item  Precision measurements of atmospheric $\nu$  parameters.
Despite the large suppression  we expect  more than 10  
$\nu _\mu C\rightarrow \mu +X$ fully reconstructed events per kTon/year.   
After the full program is
completed we should have  1\% measurement of the atmospheric oscillation
parameters.
 
\item The search for $\mutotau$ appearance will be inferred from measurements
of the neutral current rate, which would constrain the sterile neutrinos
 sector.

\end{itemize}
 
The actual program for a second phase  will depend on the outcome of the first
stage. If  $\mutoe$ appearance were discovered, then one may devote
as much as 300\,kTon/year of running time to anti-neutrinos.  (A longer exposure is
required to compensate for the smaller cross section.)
At this stage the new Proton Driver is crucial.
In this case the goals are:
\begin{itemize}
\item    Search of CP violation in the lepton sector, and  
measurement of the CP-phase.
\item  Determination  of the mass hierarchy.
\end{itemize}
If  $\mutoe$ appearance were not observed in the first stage, then the
search for $\mutoe$  would continue.

The OFF-MINOS experiment should be taking data by 2007 in 
order to be concurrent with the JHF-to-SuperK program.
Using a modular design, the OFF-MINOS  detector 
could be staged according to the availability of resources.
The  final detector design is still under discussion, but it is clear
that we need detectors that will provide high reconstruction efficiencies for charged 
$\nu_e$ current events ($\epsilon \ge 35\%$), while keeping the  
neutral current background below the two per mil level.
The detector should be located 9-10~km from the beam axis and at a
baseline between 735-900~km from the  neutrino source.

This experiment is compatible with the 
MINOS program. 

The  OFF-MINOS experiment represents an important and unique opportunity to
take full advantage of the NuMI investment and expand the neutrino oscillation 
program at Fermilab. A Proton Driver will be crucial to reach the ultimate
goal of neutrino oscillations, CP-violations.

\subsection{Neutron Program}

{\em     The new Proton Driver~\cite{pdriver} will give a proton beam in
the MW range that could produce a world-class spallation
neutron source using a small fraction~(4/30) of its total power.
An intense  room-temperature  neutron source~\cite{n2,n3}, suitable 
for a wide variety of fundamental physics experiments~\cite{n5,n6}, can be 
made  at FNAL by  optimizing the target and moderator system for long pulses 
($\Delta \tau\simeq 400 \mu$s).  This source will be complimentary to the 
short-pulse sources ($\Delta \tau < 40 \mu$s) optimized for materials 
research such as the SNS at ORNL and JHF in Japan. }\\

    The advantages of a long-pulse neutron source include~\cite{n3,n5,n6}: 
(1)~high peak intensities that can provide the most intense long-wavelength neutron
sources, (2)~the use of Time-Of-Flight~(TOF) techniques to measure energies, and 
(3)~reduced backgrounds because of the favorable duty factor. This would permit the
investigation of a number of important particle physics questions such as
time-symmetry violation, baryon number conservation, parity-nonconservation in
strong interactions, right-handed weak currents, and quantum mechanics at the
macroscopic scale~\cite{n6}. 

The primary topic, in the view of colleagues at Argonne and Fermilab,
is the measurement of the neutron electric dipole moment~(EDM).

Experimental discovery of CP violation in flavor-conserving channels would 
provide a clear indication of new physics at the electroweak scale. 
This stimulates continuing experimental efforts to detect EDM's
in elementary particles and heavy atoms \cite{n9}. 
The extraordinary precision $d_n< 6\,\times \,10^{-26}$~e-cm \cite{n10}, 
obtained in measurements of the EDM of the neutron, allows us to probe 
energy scales inaccessible in direct collider experiments.

      There have been two methods to search for the neutron Electric Dipole 
Moment. One pioneered by Norman Ramsey uses an external electric field to 
couple the EDM, and the second pioneered by Clifford Shull uses the atomic 
electric fields during Bragg scattering in a crystal.
In both cases the EDM precesses in the electric field in the same manner that 
a magnetic dipole precesses in a magnetic field and both experiments search 
for evidence of this precession using magnetic resonance techniques. 
As the amount of precession is inversely proportional to the time spent in the 
electric field, the sensitivity of the Ramsey method has been improved through 
the use of ultracold neutrons that can be stored in material bottles for 
many minutes \cite{n9,n28}. Atomic electric fields are much larger than those 
generated in the laboratory, however the time spent in the field is much 
smaller. Increasing the sensitivity of this method depends on the use of 
multiple Bragg scatters made possible by the use of perfect silicon
crystals \cite{n29}. The systematic errors differ for each method and as the 
EDM measurement is important to particle physics it is necessary to have 
confirmation of a positive result. 

      The current attempt to improve upon the Ramsey method involves a 
milli-Kelvin bath of liquid helium that effectively cools the neutrons and 
traps them \cite{n28}. The method uses neutrons of about 1~meV 
with an acceptance bandwidth of $< 1~$meV. The hope is to perform the 
magnetic resonance and detection in the same vessel that contains the liquid 
helium. Estimates made for an ILL reactor source show a possible improvement 
of three orders of magnitude in the EDM sensitivity. However, from preliminary 
tests performed at the NIST reactor \cite{n30}, it appears that backgrounds 
generated by the higher energy neutrons in the beam is a serious problem. 
One solution would be to mount the experiment at a pulsed source and use TOF 
to select only the ``useful'' neutrons. A long-pulse source will not provide the 
full 1~meV TOF resolution, but it is estimated that even a 100~meV resolution 
would provide orders of magnitude improvement in the S/N ratio. 

      In the crystal EDM experiment, thousands of successive Bragg scatters 
have been demonstrated in perfect silicon crystals~\cite{n32} suggesting that 
this experiment may also provide up to two orders of magnitude improvement in 
the EDM sensitivity. However, the experimental method requires a control 
measurement to isolate the EDM~\cite{n29}. One way to provide this would be 
to use 1.92\,A neutrons at the same time as 3.84\,A neutrons because both will 
Bragg scatter in silicon.  Each wavelength responds differently to an EDM 
precession and their effects can be separated at a pulsed source using TOF. 
 
In either case the first thing that needs to be done is to validate the model
currently under consideration to produce a spallation source starting from 8~GeV 
protons~\cite{n2}.  This first step can be made at the booster abort area, 
and a proposal is under preparation.

\subsection{P-bar Program}

{\it  As discussed in the Proton Driver physics study~\cite{booster},
the addition of the Recycler Ring to the Fermilab 
accelerator complex provides an opportunity to continue the program of  
$\bar{p}$ physics with the Anti-proton Source Accumulator. We have found
that the Proton Driver will have a negligible
impact on luminosity delivered to the collider\footnote{This statement 
is not true if there are other competitors for protons from the Main 
Injector (e.g. NuMI, CKM, etc).}, while significantly
increasing the  luminosity received by an experiment at the Accumulator. 
}\\

The operational scenario for the  $\bar{p}$ source presented here utilizes 
the Recycler 
Ring as an anti-proton bank from which the colliders makes `withdrawals' as 
needed to maintain the required luminosity in the Tevatron.  The Accumulator 
is only needed to re-supply the bank between withdrawals. When the $\bar{p}$
stacking rate is sufficiently high, and the luminosity requirements of the 
Collider experiments are sufficiently low, there will be time
between Collider fills and subsequent refilling of the recycler to deliver 
beam to an experiment in the Accumulator.  In the scenario
described in~\cite{booster}, the impact of the Accumulator experiment 
on the luminosity 
delivered to the Collider experiments is very small.  
If the Run II anti-proton stacking rate goals are met, the operational 
conditions required for running Accumulator based experiments
will be met during the BTeV era.  A simple model of the operation of the 
Fermilab accelerator complex for BTeV and an experiment in the
Collider has been developed~\cite{booster}.
The model makes predictions of the rate at which 
luminosity is delivered to BTeV and an Accumulator
experiment.  
The impact of the Proton Driver is incorporated into the model as a 
multiplicative factor that is applied to the Accumulator base stacking rate.

The model generates Collider stores and Accumulator stores according to 
the following priorities:
\begin{itemize}
\item  1st priority:      Put a store in the Tevatron for BTeV
\item  2nd priority:     Stack $\bar{p}$'s into the Recycler until there are 
                         sufficient $\bar{p}$'s for two Collider stores
\item  3rd priority       Stack and decelerate $\bar{p}$'s for the 
			Accumulator experiment
\end{itemize}
The model accumulates the integrated luminosity delivered to BTeV and to the
Accumulator experiment.  For the results reported here, the model simulated
Collider and Accumulator running over a period of 200 days.  
The model's determination of the luminosity delivered to the Accumulator
experiment is based on measurements made during the 2000 run
of Fermilab experiment E835.

Two separate model runs were done - one with and one without an enhancement
of the stacking from the Proton Driver.
The model assumes that the Proton Driver
will increase the Accumulator base stacking rate by a factor of 
three\footnote{The effect of the Proton Driver is to increase the proton
intensity on the $\bar{p}$ production target by a factor of 3 or 4.  Without
additional upgrades to the Anti-proton Source (i.e. to the stacktail momentum
cooling), the increase in protons on target will not be
translated into the same increase in $\bar{p}$ production rate.}.
The results of these runs are summarized in Table~\ref{table_ap_5}.

\begin{table}
\caption{Model Results for  $\bar{p}$ production.\label{table_ap_5}}
\begin{center}
{
\begin{tabular}{|c|c|c|}
\hline
             & No Proton Driver &	  Proton Driver Enhanced\\ \hline
BTeV Up Time &191.71 days ~~~	95.85\% &192.75 days~~~ 	96.32\%\\
Acc. Expt. Up Time&	64.41 days~~~ 	32.21\% &126.55 days~~~ 63.24\%\\
Collider Stacking &112.58 days~~~ 	56.29\% &41.56 days~~~ 	20.77\%\\
Accumulator Stacking&	7.63 days~~~ 	3.82\% &11.35 days~~~ 	5.67\%\\
BTeV  $\int Ldt$	&92.80 pb$^{-1}$/week	 &93.44 pb$^{-1}$/week\\
Acc. Expt.  $\int Ldt$&	5.84 pb$^{-1}$/week &	11.47 pb$^{-1}$/week\\
\hline
\end{tabular}
}
\end{center}
\end{table}

There are two significant findings from this analysis.  The first is that,  to
the extent that the performance of the Fermilab accelerator complex is
characterized by the above model parameters,  it is possible to run an
experiment in the Accumulator without significantly impacting the collider
program.  The second finding is  that the Proton Driver will have a negligible
impact on luminosity delivered to the collider, while significantly
increasing the  luminosity received by an Accumulator experiment. \\

\section{NuMI Based Neutrino Oscillation Program Towards a PD Upgrade}

There are several outstanding issues in the field of neutrino
oscillation measurements -- is the  three generation framework correct?  Are
there sterile neutrinos?  Are any of the mixing angles exactly 0 or
exactly $\pi/4$?  Is there CP violation in the lepton sector?  While
we would like to eventually answer all of these questions, many agree
that the next big goal beyond the current round of approved
experiments is to search for evidence of $\nu_\mu \to \nu_e$ at the
atmospheric mass splitting, which would tell us if the one completely
unmeasured mixing angle, $\theta_{13}$, is non-zero.

If $\theta_{13}$ is measured to be non-zero, then this opens up the
way for the next very interesting steps, namely, exploiting matter
effects to determine the neutrino mass hierarchy, and, assuming the
solar neutrino solution is described by LMA, a search for CP violation
in the lepton sector.  In other words, as well as being interesting in
its own right, discovery of a non-zero $\theta_{13}$ defines the next
interesting things to measure -- it is not the end of the story, merely
the beginning.

Given these assumptions, the next steps for this community are clear:  one
must first focus our efforts on finding evidence for a non-zero 
$\theta_{13}$,
but keep in mind that ultimately we will want to make much higher
precision measurements.  To understand matter effects and the
mass hierarchy we will want to measure $\nu_\mu \to \nu_e$ in both the
neutrino and anti-neutrino channels.  Finally, given the long history of
``anomalies'' seen in this field (atmospheric, solar, LSND), we think it
is important for there to be a measurement of $\theta_{13}$
at more than one baseline and preferably with more than one detector
technology.

In this section we describe in detail how an off-axis neutrino
program based on NuMI can address these issues.  It could start with NuMI
in its current design and configuration.  Once the Proton Driver becomes
available, this program would go on to test for CP violation
and to measure matter effects.   We have organized this section as follows: 
First, we briefly review the
neutrino mixing matrix and oscillation probabilities, and describe the
current knowledge of neutrino masses and mixing angles, and what the near
term series of experiments should tell us.  
In Sec.~\ref{NuMI_Off},
we describe the off-axis neutrino beam, and discuss where an off-axis 
detector~(Sec.~\ref{potential})
should be located in order to maximize its physics capabilities.
In Sec.~\ref{gen_det}, we describe one example of a fine-grained 
calorimeter which could serve as a $\nu_e$ detector, and discuss its 
reconstruction efficiencies and strategies for reducing the number
of background events in order to make a realistic assessment of the physics
capabilities of the proposed program.   
In Sec.~\ref{Physics_Sensitivity1}-\ref{Physics_Sensitivity2}, 
we discuss the physics capabilities of such
a setup, for different values of the solar mass-squared difference.

\subsection{Introduction}

The SuperKamiokande experiment~\cite{SK_atm} observes an 
angular-dependent (and energy-dependent) suppression of the atmospheric
muon-type neutrino flux, confirming with much higher precision the deficit
seen by previous experiments~\cite{osc_data_atm}. 
The best interpretation for this puzzle is that
some of the $\nu_\mu$'s transform into $\nu_\tau$'s.  
On a different front, solar neutrino 
experiments~\cite{osc_data_solar,SNO_res,SNO_nc}
have consistently measured $\nu_e$ fluxes which are significantly 
smaller than those predicted by theory~\cite{SSM}. Indeed,
recent results from the SNO experiment~\cite{SNO_nc} 
imply, at the five sigma level, that there are
active neutrinos other than $\nu_e$ ($\nu_{\mu}$ and/or $\nu_{\tau}$) 
coming from the Sun.
Finally, the LSND
Collaboration~\cite{LSND} has reported an anomalous flux of $\bar{\nu}_{e}$,
which may be interpreted as evidence for
$\bar{\nu}_{\mu}\leftrightarrow\bar{\nu}_{e}$ oscillations. This experimental
evidence has not yet been independently confirmed, but will be put to the test 
in the near future by MiniBooNE~\cite{miniboone}. 

Neutrino oscillations provide, by far, the simplest and most elegant 
solution to two out of three of these puzzles.  
Of course, to accommodate all three puzzles one would have to invoke 
even more exotic processes, since with three generations of neutrinos 
there can only be two independent mass splittings, and the three signatures
imply three very different mass splittings.  
It should be emphasized, however, 
that less standard solutions~\cite{nonstandard} 
cannot yet be discarded.

Neutrino oscillations (and other types of new physics in the neutrino sector)
can also be potentially observed in terrestrial neutrino experiments, by
studying, for example, the flux of $\bar{\nu}_e$ coming from nuclear reactors 
\cite{osc_data_reactor} or studying the $\nu_{\mu}$ flux from pion or muon 
decays~\cite{osc_data_baseline}.  The current results 
significantly constrain the neutrino oscillation parameter space,
as will be discussed in more detail in the next subsection.
If the LSND anomaly is confirmed
(and is indeed a consequence of neutrino masses and mixing), a more
complicated neutrino sector is required. This intriguing possibility 
will not be considered here, but the implications for future experiments
were discussed in reference \cite{booster}

Henceforth we will assume that active neutrino oscillations are the
solution to the solar and atmospheric neutrino results.  In this case
the standard model is augmented by at least seven (and possibly nine)
new parameters. These are three neutrino masses, three mixing angles
and one Dirac phase, which define the neutrino mixing matrix, and two
additional Majorana phases, which exist only if the neutrinos are
Majorana particles. Neutrino oscillation experiments, can probe six 
of these parameters:  two mass differences, three angles, and the Dirac 
phase.  

\subsubsection{Neutrino Oscillation Formalism}

The presence of non-zero masses for the light neutrinos introduces a leptonic
mixing matrix, $U$, analogous to the well-known CKM quark mixing matrix, 
and which we know already would be far from  diagonal. 
The matrix $U$ connects the neutrino flavor eigenstates with the mass eigenstates: 

\begin{equation}
	|\nu_\alpha\rangle = \sum_i U_{\alpha i}|\nu_i\rangle,
\end{equation}

\noindent
where $\alpha$ denotes the active neutrino flavors, $e,\ \mu$ or 
$\tau$, while $i$ runs over the mass eigenstates.  
%
It is ``traditional'' to define
the mixing angles $\theta_{12,13,23}$ in the following way:
\begin{equation}
\tan^2\theta_{12}\equiv \frac{|U_{e2}|^2}{|U_{e1}|^2},~~~
\tan^2\theta_{23}\equiv \frac{|U_{\mu3}|^2}{|U_{\tau3}|^2},~~~
\sin^2\theta_{13}\equiv |U_{e3}|^2,
\end{equation} 
while 
\begin{equation}
\Im(U_{e2}^*U_{e3}U_{\mu2}U_{\mu3}^*)\equiv 
\sin\theta_{12}\cos\theta_{12}\sin\theta_{23}\cos\theta_{23} 
\sin\theta_{13}\cos^2\theta_{13}\sin\delta 
\end{equation}
defines the CP-odd phase $\delta$. 
For Majorana neutrinos, $U$ contains two further multiplicative phase
factors, but these are invisible to oscillation phenomena.

In order to relate the mixing angles and mass-squared differences
to the parameters constrained by experiments, it is convenient to define
the neutrino masses such that $m_1^2<m^2_2$ and 
$\Delta m^2_{12}<|\Delta m^2_{13,23}|$, where 
$\Delta m^2_{ij} \equiv m_j^2 - m_i^2 $ (the data, in fact, point to 
$\Delta m^2_{12}\ll|\Delta m^2_{13,23}|$). With this definition, the ``solar
angle'' $\theta_{\odot}\simeq\theta_{12}$, while the atmospheric angle
$\theta_{\rm atm}\simeq\theta_{23}$. Furthermore, reactor experiments 
constrain $|U_{e3}|^2$. The solar mass-squared difference
$\Delta m^2_{\odot}=\Delta m^2_{12}$, while the atmospheric mass-squared
difference is $\Delta m^2_{\rm atm}=|\Delta m^2_{13}|\simeq|\Delta m^2_{23}|$.
It is important to note that $m_3^2$ can be either larger or smaller
than $m_1^2,m_2^2$.  

The oscillation probability $P(\nu_\alpha \rightarrow \nu_\beta)$ 
is given by the absolute square of the overlap of 
the observed flavor state, $|\nu_\beta\rangle$, with the time-evolved
initially-produced flavor state, $|\nu_\alpha\rangle$.  In vacuum,  
it yields the well-known result:
\begin{equation}
\begin{array}{rl}
	P(\nu_\alpha \rightarrow \nu_\beta) =&\left|\langle\nu_\beta | 
		e^{-iH_0L}|\nu_\alpha\rangle\right|^2 
	      =	\sum_{i,j} U_{\alpha i}U^*_{\beta i}U^*_{\alpha j}U_{\beta j}
		e^{-i\Delta m^2_{ij}L/2E}\\[0.1in]
	=&P_{\rm CP-even}(\nu_\alpha \rightarrow \nu_\beta) 
		+ P_{\rm CP-odd}(\nu_\alpha \rightarrow \nu_\beta) \; . \\[0.1in]
\end{array}
\end{equation}
\noindent
The CP-even and CP-odd contributions are
\begin{equation}
\begin{array}{rl}
	P_{\rm CP-even}(\nu_\alpha \rightarrow \nu_\beta) =&P_{\rm CP-even}(
		\bar{\nu}_\alpha \rightarrow \bar{\nu}_\beta)\\[0.1in]
  	=&\delta_{\alpha\beta} -4\sum_{i>j}\ Re\ (U_{\alpha i}
		U^*_{\beta i}U^*_{\alpha j}U_{\beta j})\sin^2 
		({{\Delta m^2_{ij}L}\over{4E}})\\[0.1in]
	P_{\rm CP-odd}(\nu_\alpha \rightarrow \nu_\beta) =&-P_{\rm CP-odd}(
		\bar{\nu}_\alpha \rightarrow \bar{\nu}_\beta)\\[0.1in]
        =&2\sum_{i>j}\ Im\ (U_{\alpha i}U^*_{\beta i}U^*_{\alpha j}
          U_{\beta j})\sin ({{\Delta m^2_{ij}L}\over{2E}})\\[0.1in]
\end{array}
\label{prob}
\end{equation}
such that
\beq
P(\bar\nu_\alpha \to \bar\nu_\beta)= P(\nu_\beta \to \nu_\alpha) = 
P_{\rm CP-even}(\nu_\alpha \rightarrow \nu_\beta) -
P_{\rm CP-odd}(\nu_\alpha \rightarrow \nu_\beta),
\eeq
where, by CPT invariance \footnote{if CPT invariance is broken,
as shown in \cite{nos}, all the anomalies can be naturally 
addressed without introducing sterile neutrinos}, 
$P(\nu_\alpha \to \nu_\beta) = 
P(\bar\nu_\beta \to \bar\nu_\alpha)$. 
In vacuum the CP-even and CP-odd contributions are even 
and odd, respectively, under time reversal: $\alpha \leftrightarrow \beta$.

\begin{figure}[!htb]
\vspace{1.0cm}
\centerline{\epsfxsize 12.2cm \epsffile{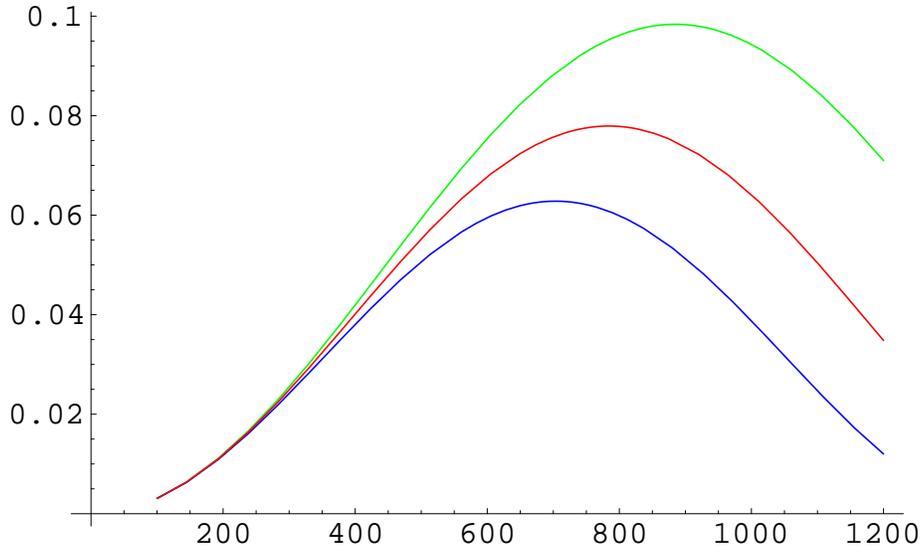}}
\caption{Transition probabilities for neutrinos (green, top curve)
and anti-neutrinos
(blue, bottom curve) in matter and vacuum  (red, middle curve)
as function of the distance for 2~GeV,  $\Delta m^2_{13}=
3\cdot 10^{-3}~{\mbox{eV}}^2$ (normal hierarchy),
$\theta_{\rm atm}=\pi/4$,
$\Delta m^2_\odot = 1\times 10^{-4}$~eV$^2$, $\theta_{\odot}=\pi/6$, $|U_{e3}|^2=0.04$,
and $\delta=0$. }
\label{uno}
\end{figure}

If the neutrinos propagate in matter, these expressions are modified, 
because of the additional forward scattering process of electron-type neutrinos
and anti-neutrinos off of electrons in matter.
The propagation of neutrinos through matter is described by the evolution
equation
\begin{equation}
i{d\nu_\alpha\over dt} = \sum_\beta \left[ \left( \sum_j U_{\alpha j} U_{\beta
j}^* {m_j^2\over 2E_\nu} \right) + {A\over 2E_\nu} \delta_{\alpha e}
\delta_{\beta e} \right] \nu_\beta \,,  
\end{equation}
where $A/(2E_\nu)$ is the amplitude for
coherent forward charged-current scattering of $\nu_e$ on electrons,
\begin{equation}
A = 2\sqrt2 G_F N_e E_\nu = 1.52 \times 10^{-4}{\rm\,eV^2} Y_e
\rho({\rm\,g/cm^3}) E({\rm\,GeV}). \,
\end{equation}
For anti-neutrinos, $A$ is replaced with $-A$, 
and $U$ with $U^*$.
Here $Y_e$ is the electron fraction and $\rho(t)$ is the matter density. 
For
neutrino trajectories through the earth's crust, the density is typically of
order 3~g/cm$^3$, and $Y_e \simeq 0.5$. 
For propagation through matter of constant density, the transition 
probabilities can be written in the form Eq.(~\ref{prob}), but by 
modifying the mass splitting and mixing angle by a constant which is a 
function of $A$, and $\Delta m_{13}^2$.  The change due to matter
effects as a function of baseline is shown in Fig.~\ref{uno}.

Long baseline (i.e. $>500km$)  neutrino experiments therefore
are sensitive to matter effects,
where the size of the effect is strongly dependent on the
baseline length and neutrino energy.  
Some unknowns related to the neutrino mass pattern can be addressed
with the ``help'' of the matter effects. As alluded to before,
the current data leave us with two alternatives for the spectrum of
the three active neutrino species: a ``normal'' neutrino mass
hierarchy or an ``inverted'' neutrino mass hierarchy.
In the case of a ``normal'' mass hierarchy, the ``solar pair'' of states is lighter
than $\nu_3$, \ie\ $m_3 > m_2, m_1$. In the case of inverted hierarchy, 
the states of the solar pair are heavier than  $\nu_3$, \ie\ $m_3 < m_2
\simeq  m_1$.
The key difference between these two hierarchies is then that, in the normal
hierarchy, the small $U_{e3}$ admixture of $\nu_e$ is in the heaviest
state whereas in the inverted hierarchy, this admixture is in the 
lightest state. 
The difference between both schemes is parameterized by the  
sign of $\Delta m^{2}_{23}$.\footnote{Another way of 
treating the neutrino mass hierarchy
is by defining $m^2_3>m^2_2>m_1^2$, and redefining the solar, atmospheric and 
reactor angle depending on whether $\Delta m^2_{12}$ is larger or smaller than 
$\Delta m^2_{23}$. In such a scheme, the reactor data would limit $|U_{e3}|^2$ 
(normal hierarchy) or $|U_{e1}|^2$ (inverted hierarchy).} 
A positive $\Delta m^{2}_{23}$  is defined as a normal hierarchy.  

In going from $\nu$ to $\overline{\nu}$, there are matter-induced CP-
and CPT- odd effects associated with the change $A \rightarrow -
A$. The additional change U $\rightarrow$ U$^*$ introduces further effects
(this is the ``genuine'' CP-violation), which are usually subleading. 
Note that the matter effects depend on the
interference between the different flavors and on the relative sign
between $A$ and $\Delta m^{2}_{23}$. As a consequence, an experimental
distinction between the propagation of $\nu$ and $\overline{\nu}$ (the
sign of $A$) can determine the sign of $\Delta m^{2}_{23}$.

The Standard Model can be extended to incorporate neutrino
masses in a variety of ways. Many theories beyond the minimal
model also lead to non standard neutrino interactions~\cite{nsni}. These
include most models of generating neutrino masses, such as
the simplest seesaw type schemes, supergravity $SO(10)$
unified theories, models of low energy supersymmetry with
broken R-parity as well as some radioactive models of neutrino
masses.

Long baseline neutrino experiments have a reasonable potential
for probing non standard neutrino matter interactions as
well as some kinds of new physics,
specially if a non zero CP violating effect is detected.
The role of a long baseline experiment in testing this 
kind of new physics is complementary to efforts to probe
similar flavour changing effects in the charged lepton sector
while being model independent.

CP violating observables are particularly sensitive
to new physics because they are not necessarily  as suppressed
by the small mass differences and mixing angles as the
standard contributions. Even more, the dependence
on the distance between the source and the detector is in
general different, and this might help in disentangling
standard effects from the new ones.

If one tries to incorporate the LSND signal
only using oscillations between active flavors, 
all data can be consistently fit
by incorporating CPT violation~\cite{nos}. CPT violation,
unlike CP or T violation, can also be detected in survival
probability measurements, giving long baseline experiments
the chance to discover it or set stringent limits on this possibility.

\subsubsection{Present Status of the Oscillation Parameters}


Many analyses of the solar, atmospheric, and reactor neutrino data can
be found in the literature, including two-flavor and three-flavor
analyses of the solar data~\cite{osc_anal_solar}, two-flavor analyses
of the atmospheric data~\cite{osc_anal_atm}, three-flavor analyses of
the combined atmospheric and reactor data~\cite{osc_anal_atm_reac} and
combined analyses of all neutrino data~\cite{osc_anal_comb}.  It
should be noted that the experimentally allowed range for the
oscillation parameters varies depending on a number of assumptions:
which data are taken into account, how many neutrino species
participate in the oscillation, what was the statistical ``recipe''
used to define allowed regions, etc.  Here we will summarize the
current ``standard'' results.

For the ``atmospheric and reactor parameters'' one obtains 
at the 99\% confidence level (CL)~\cite{osc_anal_comb},
\begin{eqnarray*}
&|U_{e3}|^2<0.06, \\
&0.4<\tan^2\theta_{\rm atm}<2.5, \\
&1.2\times 10^{-3}~{\rm eV}^2<|\Delta m^2_{23}|<6.3\times 10^{-3}~{\rm eV}^2.
\end{eqnarray*}

The situation of the ``solar parameters'' is far less certain. 
There are different disjointed regions of the parameter space which satisfy the current
solar neutrino data. They are traditionally referred to 
as: SMA (LMA), the small (large) mixing angle MSW solution,
LOW, the once low probability, now low $\Delta m^2_{12}$ MSW solution, and the 
various ``vacuum solutions'' are called VAC. 
Of the four regions, two (LMA, and LOW) are very robust, and appear in 
different ``types'' of data analysis. The VAC solutions are rather unstable, and can 
``disappear'' if the data is analyzed in different fashions. The SMA 
solution is currently ruled out at more than the three-sigma level, 
but should not be completely discarded yet. Numerically, at the 99\% CL (according to
Bachall {\it et al.}\/ in~\cite{osc_anal_solar}),
\begin{eqnarray*}
&0.25<\tan^2\theta_{\rm atm}<0.8, \\
&2.1\times 10^{-5}~{\rm eV}^2<\Delta m^2_{12}<2.7\times 10^{-4}~{\rm eV}^2, \\
&{\rm or}\nonumber \\
&0.5<\tan^2\theta_{\rm atm}<0.7, \\
&6\times 10^{-8}~{\rm eV}^2<\Delta m^2_{12}<1\times 10^{-7}~{\rm eV}^2, \\
&{\rm or}\nonumber \\
&1/3<\tan^2\theta_{\rm atm}<3, \\
&4\times 10^{-10}~{\rm eV}^2<\Delta m^2_{12}<7\times 10^{-10}~{\rm eV}^2, 
\end{eqnarray*}

In summary, while some of the oscillation parameters are know with some ``precision''
(the atmospheric mass-squared difference is known within a factor of roughly six),
the information regarding other parameters is very uncertain. In particular, 
$\tan^2\theta_{\odot}$ can be either very small ($\sim 10^{-4}-10^{-3}$), 
or close to one, while $\Delta m^2_{12}$ can take many different values, from around
$10^{-10}$~eV$^2$ to more than $10^{-4}$~eV$^2$. Finally, there 
is absolutely no information on the CP-violating phase $\delta$, or on the sign
of $\Delta m^2_{23}$, while for $|U_{e3}|^2~(\theta_{13})$ only a moderate upper bound has been 
established.
 
\subsubsection{Prospects}

The precision with which some neutrino oscillation parameters are
known will improve significantly in the near future, and it is almost
certain that the ambiguity in the solution to the solar neutrino
puzzle will disappear.

The values of $|\Delta m^2_{23}|$ and $\tan^2\theta_{\rm atm}$ should be better
determined by long-baseline neutrino experiments~\cite{K2K, MINOS, CNGS}. 
In particular, the MINOS experiment~\cite{MINOS} expects to 
measure these atmospheric
parameters with order 10\% uncertainties, as does the 
CNGS program~\cite{CNGS}. The sensitivity to $|U_{e3}|^2~(\theta_{13})$, on the
other hand, is supposed to be limited to at most a few percent (for example, 
close to three sigma excess of $\nu_e$ events can be obtained after four years of 
ICARUS running for $|U_{e3}|^2=0.01$ \cite{CNGS}). The K2K experiment started taking
data in~1999 (and is to resume data-taking by the end of the year), while the NuMI
(CNGS) project is scheduled to begin in early 2005 (2006). 
    
In the solar sector, different solutions will be explored by different analyses
of data from different experiments. 
The LMA solution to the solar neutrino puzzle will be either established or excluded
by the KamLAND reactor experiment \cite{KamLAND}. Furthermore, if LMA happens to be
the correct solution, KamLAND should be able to measure the oscillation parameters
$\tan^2\theta_{\odot}$ and $\Delta m^2_{12}$ with good precision by analyzing the
$\bar{\nu}_e$ energy spectrum, as has been recently
investigated by different groups \cite{kamland_results,kamland_plus_solar,prospects_summary}. 
Three years of KamLAND running
should allow one to determine, at the three sigma level, 
$\Delta m^2_{12}$ within 5\% and $\sin^2 2\theta_{\odot}$ within 0.1. A combination of KamLAND
reactor data and solar data should start to address the issue of whether 
$\theta_{\odot}$ is smaller or greater than $\pi/4$ \cite{kamland_plus_solar}. 
The KamLAND experiment ``turned on'' early in 2002, and should already have results by the
end of 2002 or early in 2003 \cite{kamland_new}.

The LOW solution will be either excluded or unambiguously established \cite{day-night}
by the Borexino experiment \cite{borexino} 
(and by a possible upgrade of the KamLAND experiment, such that it can be
used to see $^7$Be solar neutrinos). This is due to the fact that,
if the LOW solution is correct, the $^7$Be solar neutrino flux should vary dramatically
as a function of zenith angle. In particular, an analysis of the zenith angle
dependency of the Borexino data should allow one to measure, at the three sigma level,
$\Delta m^2_{12}$ within a factor of three (say, in the range 1 to $3\times 
10^{-7}$~eV$^2$) and $\tan^2\theta_{\odot}$ within 0.2 \cite{day-night} (see also 
\cite{prospects_summary}). 
These estimates are very conservative and do not depend, for example, on 
the solar model prediction for the $^7$Be neutrino flux \cite{day-night}. 
 
Solutions with $\Delta m^2_{12}$ less than a few $\times 10^{-9}$~eV$^2$ and greater 
than a few $\times 10^{-11}$~eV$^2$, and $\tan^2\theta_{\odot}$ between roughly 
0.01 and 100 will also be either excluded or established by experiments capable 
of measuring the $^7$Be solar neutrino flux. It turns out that in this region of 
the parameter space the flux of $^7$Be solar neutrinos depends very strongly on 
the Earth--Sun distance, and anomalous seasonal variations should be readily observed, 
for example, at Borexino. Estimates of the performance of Borexino (and KamLAND) data 
obtained in \cite{seasonal} indicate that, even if very conservative assumptions are
made (they do not rely, for example, on assuming that the $^7$Be solar neutrino flux 
is known), $\Delta m^2_{\odot}$ can be measured at better than the percent level (see
also \cite{prospects_summary}). The Borexino experiment is supposed to start taking
data later in 2002.

Other less definitive possibilities are still available. The SNO experiment 
experiment may eventually provide enough information for resolving the ambiguities 
in the solar neutrino sector. A significant amount of research effort has been 
devoted to this issue \cite{SNO_spectrum,SNO_generic}. Further information may also
be obtained if neutrinos from a nearby supernova are detected~\cite{supernova}.

Finally, it is important to mention that non-oscillation experiments can also
contribute to the understanding of neutrino masses and leptonic mixing angles.
In particular, future searches for neutrinoless double beta decay~\cite{doublebeta_exp}
are not only capable of measuring a particular combination of the Majorana 
neutrino phases, but can also help piece together  the solar neutrino puzzle 
\cite{doublebeta_theo}.
   
In summary, this and the next generation of neutrino experiments have
the potential to establish neutrino oscillations (or at least neutrino
flavor conversions) and to measure roughly 
most of the leptonic mixing angles and
mass-squared differences. In particular, it seems very plausible that
the larger mass-squared difference and the atmospheric mixing angle
will be measured to about 10\%, while the small mass-squared
difference and the solar mixing angle will be known with a precision
better than one order of magnitude (in the case of the LMA solution,
at around the 10\% level).

The absolute value of the $U_{e3}$ element of the neutrino mixing
matrix may or may not be measured in the next round of experiments, but it 
is clearly the key to understanding the remaining outstanding issues: 
is there CP violation in the lepton sector, and what is the mass 
hierarchy?.  The sensitivity of the current round of 
long baseline accelerator experiments is within less than one
order of magnitude of the current reactor bound.  The proposed long-baseline
neutrino experiment (phase one) from the future JHF facility to
SuperKamiokande~\cite{JHF} should be able to improve on the limit on
$|U_{e3}|^2$ by at least one order of magnitude ($|U_{e3}|^2<0.0015$
at the 90\% CL), but because it is a shorter baseline, would not be sensitive
to matter effects.  


One might ask, why is getting farther in reach for $|U_{e3}|^2$ such a daunting
talk?  The answer is simply that the technique accelerator experiments have at 
hand is simply to look for $\nu_e$ appearance in a $\nu_\mu$ beam, which already
has some intrinsic $\nu_e$ contamination.  Furthermore, it turns out that 
neutrino interactions, in particular, neutral current interactions, are much 
more easily misidentified as $\nu_e$ charged current events, than as $\nu_\mu$ 
charged current events.  Both the three-body decays which produce $\nu_e$'s, and 
the neutral current interaction provide backgrounds that are very broad in 
reconstructed energy, so by using a mono-chromatic $\nu_\mu$ beam, and a detector
with correspondingly good energy resolution, one can significantly 
reduce both sources of backgrounds.  


In the next section we describe the beamline that is in fact already being constructed
for the MINOS experiment, and how that very same beamline can provide a very clean
almost mono-energetic neutrino beam at different locations, remote from where the 
MINOS detector now stands.  In the following section we describe the different 
possible detector technologies that could be used for one of these new beams, 
and finally after that we give a specific example of one detector concept, how it 
can further reduce backgrounds, and finally, the physics reach of that kind of 
detector, both without and then with a proton driver upgrade.  

\subsection{NuMI Off-Axis Beams}
\label{NuMI_Off}

The Neutrinos at the Main Injector~(NuMI)~\cite{NuMI} beamline was 
designed to provide an intense $\nu_{\mu}$ beam to the MINOS 
experiment~\cite{MINOS}by impinging  120~GeV protons on  a graphite target.
The $\nu_{\mu} (\bar{\nu}_{\mu})$'s are derived mostly from  
secondary $\pi^+ (\pi^-)$ decays, with kaons contributing significantly only above 
10-15 GeV.  
The MINOS detector, in the Soudan mine, is located at a distance of 735~km from 
FNAL, and the beam line is built to point directly at the MINOS detector.  This 
configuration gives the largest total number of events, and a correspondingly 
broad energy distribution, which is very important to establish what the mechanism
is for neutrino disappearance observed in SuperKamiokande.  However, if one assumes
that the mechanism is in fact oscillations, then as mentioned earlier, 
the next important step in the field is to search for $\nu_\mu \to \nu_e$ appearance
at the same mass splitting.  By placing detectors 
at different locations an experiment would be able to use very different
fluxes to make measurements, and it turns out that these fluxes are much 
better suited to take this next step than an on-axis beam, because of the 
very narrow energy distribution of the resulting $\nu_\mu$ beam. 
 
\subsubsection{Neutrino Fluxes} 
\begin{figure}[htb]
\centerline{\psfig{file=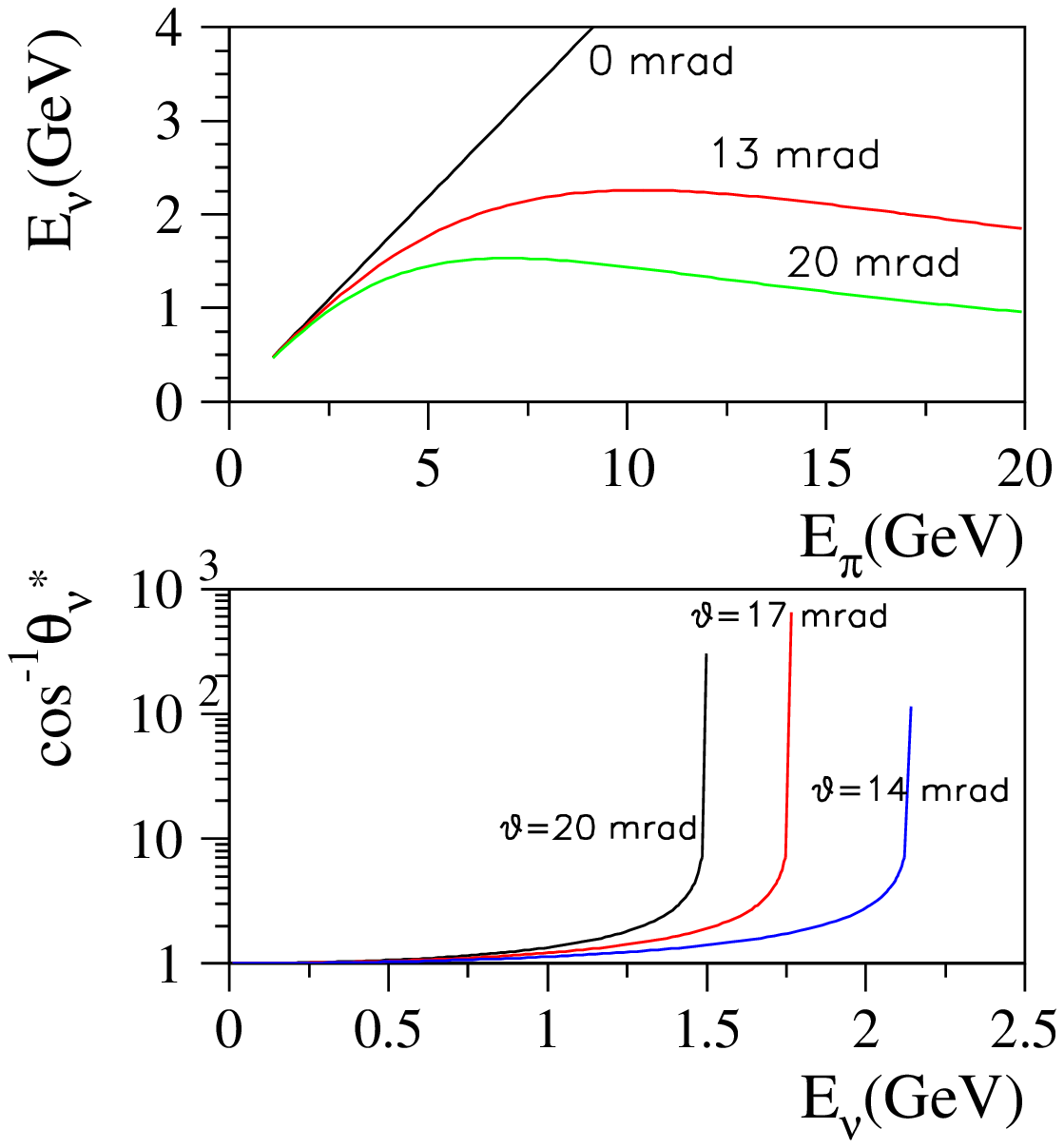,width=10cm} 
\psfig{file=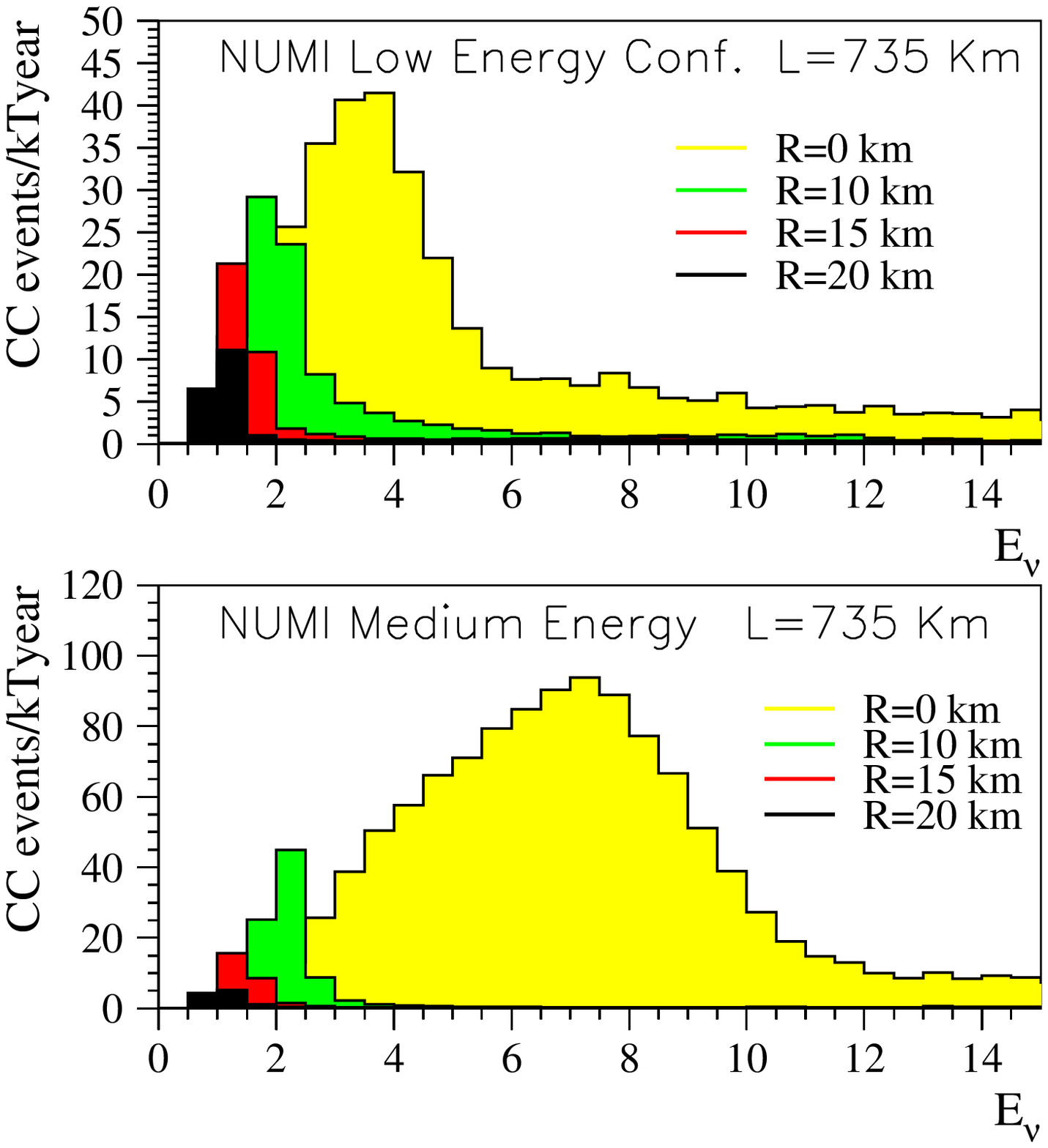,width=9cm}}
\caption[0]{\label{off_theory}
Top-Left -- allowed $E_\nu$ values for several $\theta_\nu$ values. 
Bottom-Left -- shows  $1/\cos\theta^*_\nu$ that is proportional to 
the neutrino flux. 
$\cos\theta^*_\nu=\sqrt{1-\frac{E^2_\nu}{E^{*2}_\nu}\tan^2\theta_\nu}$   
is the $\nu$ production angle in the rest frame of the $\pi$.
Right -- On and off-axis beams for the low and medium
energy NuMI horns configuration.  A full beam simulation  was made using the 
Geant based GNuMI Monte Carlo. 
}
\end{figure}

The neutrino beam energy spectra at any location can be 
predicted from energy and momentum conservation in the $\pi$ decay process:
\begin{equation}
\label{cons}
E_\nu=\frac{m^2_\pi-m^2_\mu}{2(E_\pi-p_\pi \cos\theta_\nu)}  
     =\frac{0.004~\rm GeV}{(E_\pi-p_\pi \cos\theta_\nu)}, 
\end{equation}
where $m_\pi$ and  $m_\mu$ are the pion and muon 
rest  masses, $E_\pi$ and $p_\pi$ are the 
pion energy and momentum, and $\theta_\nu$ is the angle at which the 
neutrino is emitted with respect to the pion direction.  The  maximum angle 
in the lab frame  relative to the  pion direction is related to the 
neutrino energy by:
\begin{equation}
\label{cons_max}
\theta^{max}_\nu = \frac{(30 +\Delta_T)~\rm MeV}{ E_\nu}, 
\end{equation}
where 30~MeV is the neutrino momentum in the rest frame of the pion, 
and   $\Delta_T$ keep into account the nonzero transverse momentum of the 
decaying $\pi$.
As shown in 
Fig.~\ref{off_theory}(a),  if $\theta_\nu\simeq 0$ the neutrino energy 
is proportional to the pion energy ($E_\nu=0.44E_\pi$), while at an off-axis 
location ($\theta_\nu\neq 0$) there is a maximum neutrino energy which is 
independent of the energy of the parent pion. Therefore,  the off-axis 
configuration allows one to use a fraction of the ``total'' beam that is 
characterized by having lower $E_\nu$. The maximum flux for a fixed $E_\nu$
will be obtained  when operating close to the corresponding 
$\theta^{max}_\nu$, see Fig.~\ref{off_theory}(b). The lower energy neutrinos 
provided by NuMI off-axis beams are highly desirable because they allow 
beams which are more suitable for studying $\nu_\mu \to \nu_e$ transitions, 
which, have yet to be seen at long baselines.

Two toroidal magnetic horns sign and momentum-select the secondary 
beam.  The horns are movable allowing one to obtain different neutrino energy 
spectra.  For example, Fig.~\ref{off_theory} depicts the expected energy 
spectrum at different location for the low energy (horns 10~m apart) and  
the medium energy (horns 27~m apart) horn configurations. 
As shown, the off-axis beams are characterized by having a 
narrow and well-defined energy distribution with fluxes   higher than the 
corresponding on-axis energy. In addition, the harmful high 
energy tail~(a source of NC background) of the on-axis beams is not present, as expected from energy and 
momentum conservation, Eq.~(\ref{cons}).
All these  distributions are calculated using the GNuMI Geant based 
Monte Carlo~\cite{GNuMI}, and 
include the full beamline, target and decay pipe description.


A reduction of the high energy tail of the medium energy  off-axis beam is 
in  Fig.~\ref{off_theory}. There are two reasons for this: (1) the 
transverse momentum of the 
pions is smaller for the medium energy configuration  making 
$\Delta^{medium}_T < \Delta^{low}_T$, and (2) the mean energy of the kaons
in the medium energy beam is higher producing neutrinos that are of higher 
energy and therefore less harmful to the analysis.

\subsubsection{Anti-neutrino Fluxes} 

\begin{figure}
\centerline{\psfig{file=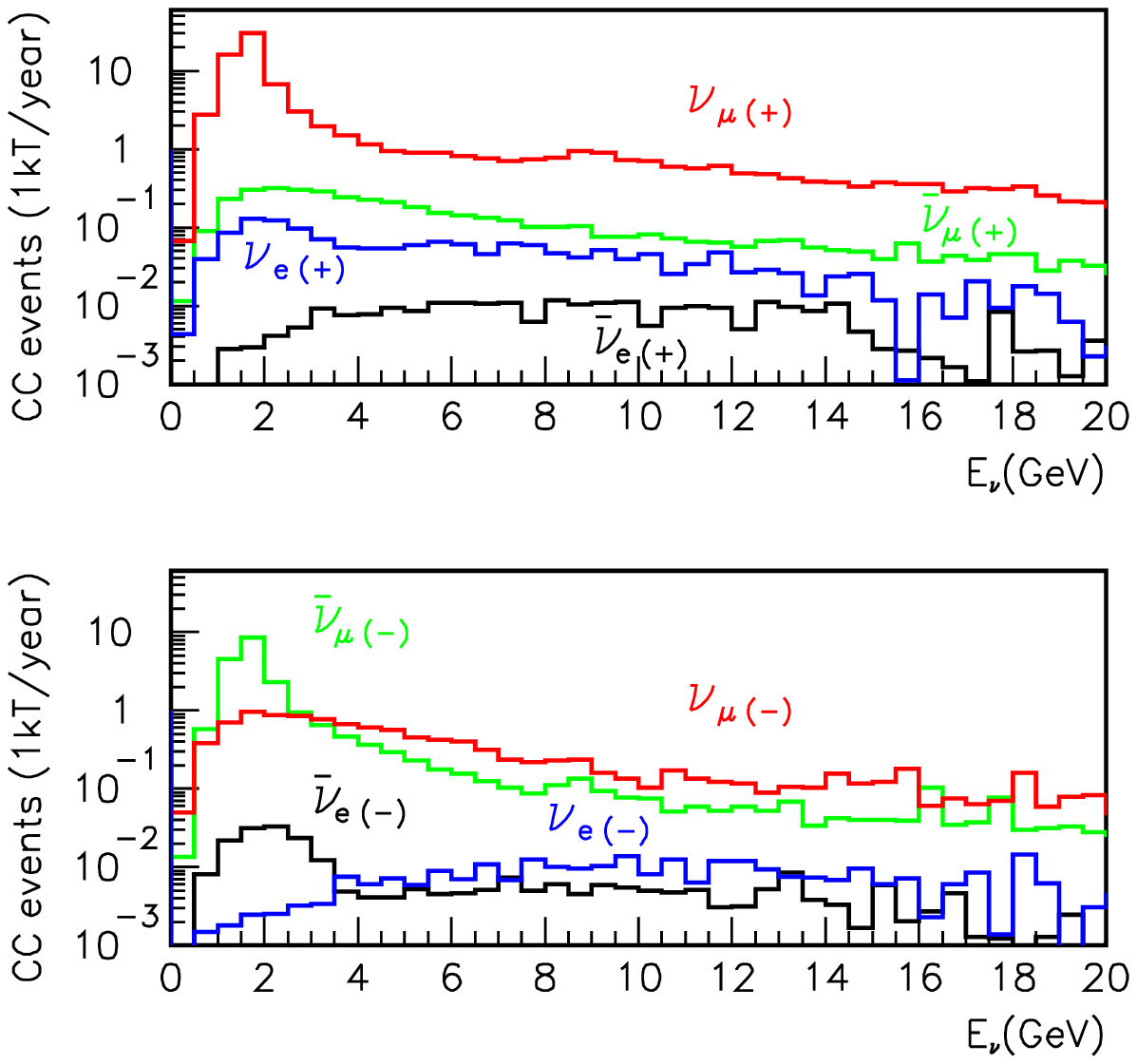,width=7.cm}
\psfig{file=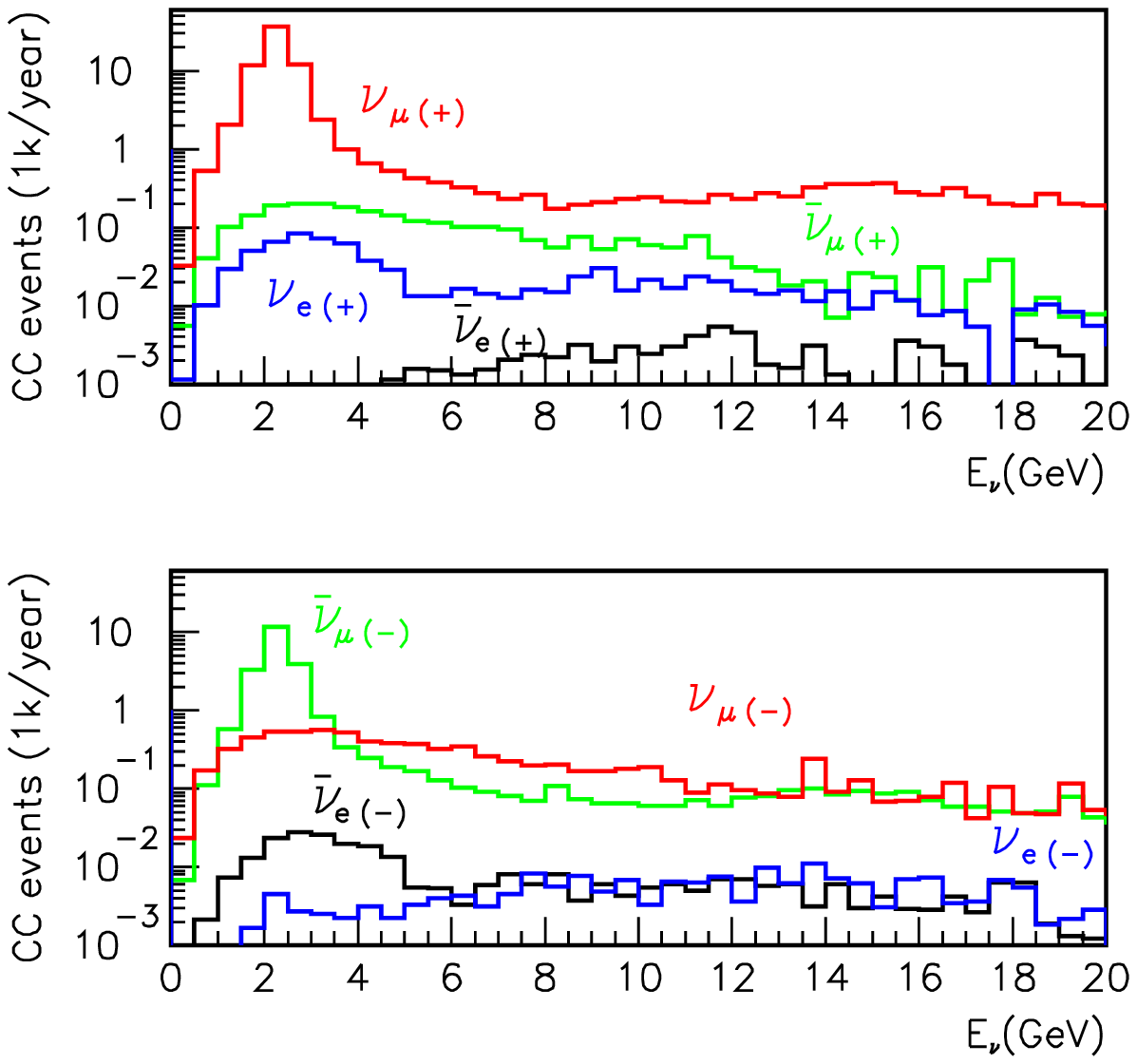,width=7.cm}}
\caption[0]{\label{beam_comp} Beam composition for positive (+)
and negative (-) horn currents. 
left: for the a low energy configuration 
at $L=735$~km and $R=10$~km.
right: for the a medium  energy configuration 
at $L=900$~km and $R=12$~km.
}
\end{figure}

Studies of matter effects and CP violation require a comparison between 
$\nu_\mu$ and  $\bar{\nu}_\mu$ oscillations. An anti-neutrino beam can be
produced simply by reversing the polarity of the horns.  This beam's flavor 
composition and $(\nu_e+\bar{\nu}_e)/(\nu_\mu+\bar{\nu}_\mu)$ 
are also shown in Fig.~\ref{beam_comp}.  Although the $\nu_\mu$ contamination 
in the $\bar\nu_\mu$ beam is significantly larger than in the $\nu_\mu$ beam, 
the fractional $\nu_e+\bar\nu_e$ contamination 
in the $\bar\nu_\mu$ beam is roughly the same as in the $\nu_\mu$ beam, 
and in fact the ratio of the 
event rates in a far detector is primarily due to the difference in neutrino 
and anti-neutrino cross sections.  At 2~GeV the cross section 
ratio is about 3, so to measure
the same probability to equal precision in anti-neutrino running would take 
approximately three times longer in $\bar\nu$ studies.  However, between matter
effects and CP Violation, the $\bar\nu_\mu \to \bar\nu_e$ 
oscillation probability might be significantly
larger than its CP conjugate, so in fact the required running time to extract 
physics from $\bar\nu$'s depends very much on what those parameters turn out to be.  
In later sections of this report we give sensitivities to physics parameters assuming 
300\,kTon -years of $\bar\nu$ running compared to 120\,kTon -years of $\nu$ running, 
but certainly the length of time needed to see physics in $\bar\nu$ running 
depends on what the $\nu$ running has measured.  

\subsubsection{Systematic Uncertainties}  
\label{beam_mon}

The beamline and resulting neutrino beams described earlier in this
section can (and should!) be used for both $\nu_e$ appearance and
$\nu_\mu$ disappearance measurements.  The two different analyses,
however, depend very differently on systematic uncertainties.  In this 
section we describe how systematic effects will contribute to the two 
analyses, and how the presence of both the MINOS on-axis near detector, 
and a possible on or off-axis near detector of the same technology as 
the far off-axis detector can be used.  

For both measurements, however, one must be able to predict the
non-oscillated $\nu_\mu$ spectrum to high accuracy.  The MINOS
experiment will have an on axis near detector which is functionally
identical to the on axis far detector, and should be able to measure
the $\nu_\mu$ event rate for the on-axis beam. As long as the observed
neutrinos come from one and the same parent pion (or kaon) beam, the
unoscillated $\nu_\mu$ spectra at two arbitrary detector locations are
always strongly correlated and a measurement of one of them makes
possible a prediction of the other.  The technique of doing so has
been shown in Ref.~\cite{michal_adam1}, and involves a correlation
matrix: for every $\nu_\mu$ event at a given energy reconstructed in
the near detector, there is a collection of $\nu_\mu$ events expected
at various energies in the far detector.  Because the events in the
peak of both the on-axis and off-axis beams come from pion decays, the
uncertainty in the $\nu_\mu$ flux due to hadron production alone 
can likely be reduced do 1-2\% in the peak of the off-axis energy distribution.

     An outstanding issue is the one of cross sections.  With neutrino
energies being significantly different in the near and far sites,
uncertainties from this source will not cancel in the
prediction.  Ideally, precision measurements of neutrino cross
sections would be required.  It also possible to infer those cross
sections indirectly as part of the same project, once hadron production
is precisely measured.

     In the absence of oscillations, one might 
expect roughly 15 accepted events/kTon-year 
between 1.5 and 2.5\,GeV for an off-axis beam.  So for a 100\,kTon-year run, if 
$\theta_{23}$ really was $\pi/4$, then the statistical error on the oscillation 
probability could be as low as 1 to 3 divided by 1500, which is well below 
the 1-2\% quoted above.  So obviously the systematic 
uncertainty on the flux and the acceptance will be the dominating effect, and 
a near off-axis detector (combined with improved hadron production measurements, for 
example at E907) may improve this.  Comparing the relative 
efficiency of an off-axis detector
with the MINOS on-axis near detector to a few per cent would also be a challenge, 
which again would stress the need for a near detector of the same technology
as the far detector.  

By contrast, in an appearance search, 
the systematic uncertainties could actually be small compared to the 
uncertainty due to statistical fluctuations in the background.  So, for example, 
if one expects roughly 0.3 intrinsic $\nu_e$ background events accepted per kTon year 
the statistical uncertainty due to the expected number of background events
in a 100kTon-year experiment would be $1/\sqrt{30}$, or about 20\%.  Therefore, 
a 10\% systematic uncertainty on the intrinsic $\nu_e$ flux would have negligible
effect on the physics sensitivity.  For $\nu_e$ background events in the signal 
region, most of them arise from muon decays, which in turn arise from the same 
parent pion decays that have been measured in the peak of the neutrino energy
distribution in the MINOS near detector.  The small fraction which arise from kaon 
decays can be constrained again by E907 measurements, and from extrapolating from 
higher energy $\nu_e$ events in the far detector.

So, while 10\% uncertainties in the $\nu_e$/$\nu_\mu$  flux ratio 
are acceptable and even attainable using the MINOS on axis near detector, 
the neutral current backgrounds may prove to be far more challenging.  
If the neutral current background is half the size of the intrinsic $\nu_e$ 
background, then the systematic uncertainty 
could be twice as large (i.e. 20\%) and still be negligible.  On the other hand, if the
neutral current background is twice the intrinsic $\nu_e$ background, then it
needs to be known twice as well, or to 5\%.  There are currently large uncertainties
on the neutral current cross section in the first place, so further studies 
(and measurements!) are needed to demonstrate how precisely one would need to 
know the neutral current background.  These measurements would best be made in a 
dedicated off-axis detector of the same or more segmentation somewhere 
in the NUMI facility (possibly in the access shaft).

\subsection{Potential  Off-Axis Experiments}
\label{potential}

     There are several possibilities for a detector for the NuMI off-axis 
that are being considered~\cite{LOI}.  One can summarize the detector's performance in 
this beam by two numbers, namely the efficiency for detecting a 2~GeV electron 
neutrino charged current event, and the efficiency for removing any neutrino 
interaction that is not a electron neutrino charged current event.  There 
will be both $\nu_\mu$ charged current and $\nu_\mu,\nu_\tau$, and $\nu_e$ 
neutral current interactions in this detector, and certainly the neutral 
current interactions are those which can most easily fake a $\nu_e$ charged
current event.  Of course, the physics reach for a given detector (and for 
a given cost) depends on those two performance numbers, as well as a 
cost per kTon of detector.  We will first discuss the physics performance.  

     Since the intrinsic $\nu_e$ content in the NuMI off-axis beam is roughly 
$0.5\%$ when integrating from 1.5 to 2.5~GeV, the goal for any detector 
is to reduce the NC background to approximately this level.  There are roughly 
three kinds of detectors which are being considered: fine-grained calorimetry
(a absorber/readout sandwich with different options for both absorber and 
readout), a water cerenkov device   like SuperKamiokande, 
and a Liquid Argon TPC, like ICARUS.  

Because a liquid Argon TPC would have
by far the finest segmentation and resolution, it could 
presumably have the highest detector efficiency and lowest background.  By 
cutting on the dE/dx of the electron candidate track in the first radiation 
length, one can remove essentially all of 
the neutral current events, while retaining 
90\% of the signal~\cite{ICARUS}. 

Next in segmentation is the fine-grained calorimeter concept.  
Studies of various absorber and readout materials produce slightly different 
efficiencies, but roughly speaking, with a certain set of cuts, 
the NC events represent a background 
which is about half of the intrinsic $\nu_e$ background, while retaining
roughly 40\% signal efficiency~\cite{offaxis,michal_ni}.

Finally, there is water cerenkov technology, which 
so far has been demonstrated in simulations (using SuperK-based 
phototube coverage, electronics, noise, etc.) to reduce the NC background to 
approximately twice the size of the intrinsic $\nu_e$ background, but 
with a signal efficiency of about 25\%, because only quasi-elastic events 
are used in the analysis~\cite{mark_ni}.

\begin{figure}[here]
\begin{center}
\epsfig{file=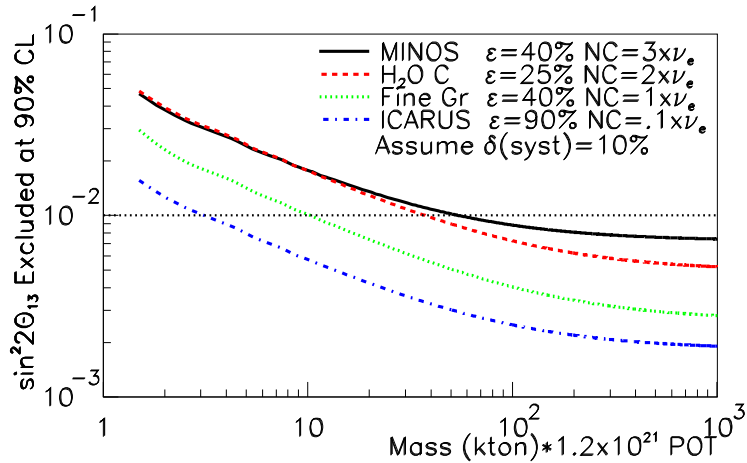,width=10.cm}
\end{center}
\caption{\label{detector1}
Expected detector performance.
}
\end{figure}

All of these studies are preliminary and work is continuing to optimize
the detector responses, but for the moment one can consider in general
three types of detectors:  the next questions to ask are: how many kTons are 
needed for the three different technologies to provide comparable physics
reach, and how good must the systematic uncertainty be on the background
rejection?  The answer depends very much on what reach one is aiming for:
for a factor of 10 past the CHOOZ limit, systematics at the 20\% level are 
acceptable, and roughly speaking, 100\,kTon-NuMI-years of fine-grained 
calorimeter has the same reach as 25\,kTon-NuMI-years 
of liquid argon TPC, and 400\,kTon-NuMI-years of water cerenkov.  
This can be shown in Fig.~\ref{detector1}.  However, for a factor of 
100 past the CHOOZ limit, the systematics become more important, and 
because of the higher background levels currently attained, Water cerenkov 
hits systematic
limitations much sooner than the other detector technologies.  It should 
be noted, however, that if Water Cerenkov can be shown to have better 
background rejection, it may remain attractive, because since only the 
surface need be instrumented, the cost scales like $mass^{2/3}$.

While the physics performance is very different 
between the three,  the cost/kTon is also quite different.  In fact the 
cost/kTon is perhaps an even greater unknown than the physics performance.  
For example, if a 5kTon liquid argon calorimeter were to be built using 
600~Ton modular structures like those used in ICARUS, the cost would 
be prohibitively high.  If it were built using a single 7\,kTon(total) module, 
a miniature version of the LANNDD detector~\cite{Clinestalk} then it would 
be considerably cheaper, perhaps as low as $40$ Million 
dollars~\cite{Clinestalk}.  
A 20\,kTon fine-grained calorimeter has been 
costed as low as 30-60~Million dollars, and finally, a Water cerenkov at 
80\,kTon has been costed at 85~M/kTon for four water tanks~\cite{milind2}
It is still too preliminary to make a detector choice now as 
these numbers as well as some of the performance numbers are quite 
preliminary.  What is very clear between all of these detectors, however, 
is the fact that the cost of a Proton Driver is a far more cost-effective 
way to extend the physics reach, compared to increasing the detector mass.  
It should be noted that five times any detector cost proposed would be 
considerably more expensive than building a proton driver upgrade.

\subsection{$\nu_\mu$ and $\nu_e$ Detector Simulation}
\label{gen_det}

This section presents possible analyses towards $\nu_e$ appearance
and $\nu_\mu$ disappearance in a highly segmented calorimetric detector,
based on simulated data.  As an example, in the following studies
we have assumed a detector made up of iron foils interleaved with 1\,cm
thick scintillator planes.
The scintillator strips are oriented at plus or minus $45^\circ$
from vertical, alternating every other plane (two views).
Detector performance was studied for different longitudinal and
transverse segmentations, as well as for different spacings between
two adjacent planes.  Unless stated otherwise, throughout most of this
section we will be assuming a longitudinal
sampling every quarter of a radiation length (which in case of iron
means a thickness of 4.5\,mm per foil), the transverse width
of each readout cell is chosen to be 2\,cm, and a 3\,cm long
air gap is left between
each two iron-scintillator pairs.

\subsubsection{GMINOS Implementation}

All the detector simulations were carried out with the aid of the
GMINOS program, developed at Fermilab as part of the official fortran-based
code for the MINOS experiment.  GMINOS is a Geant-based Monte Carlo
program which was widely used during the design period of the MINOS
detectors.  It simulates particle interactions and detector responses
in a largely user-defined detector, leaving a lot of freedom for the
definition of detector geometry, choice of passive and active materials
and readout techniques, and of the kind of reactions under study, without
the need of writing additional code or recompiling the program.  The
actual geometry of the detector is defined via user input cards, under
the assumption of the most general modular detector structure: a
repeating pattern of passive absorption planes and active detector
planes oriented at some fixed angles with respect to the beam axis.
Plane definition includes details of the internal structures surrounding
readout cells/strips to account for the basic construction features of
of fiber-scintillator, RPC and LST detectors.  Each volume element is of
a user-specified material and dimensions.

Neutrino interactions are generated using the NEUGEN cross sections
and event generator.  All hits coming from particles that deposit energy
are recorded, the record contains information about particle identity,
its momentum, energy loss and time-of-flight for each active volume
being hit.  Electronics noise and inefficiencies are included,
attenuation effects in light propagation simulated and corrected for
during event reconstruction.

\begin{figure}[htb]
\begin{center}
\epsfig{file=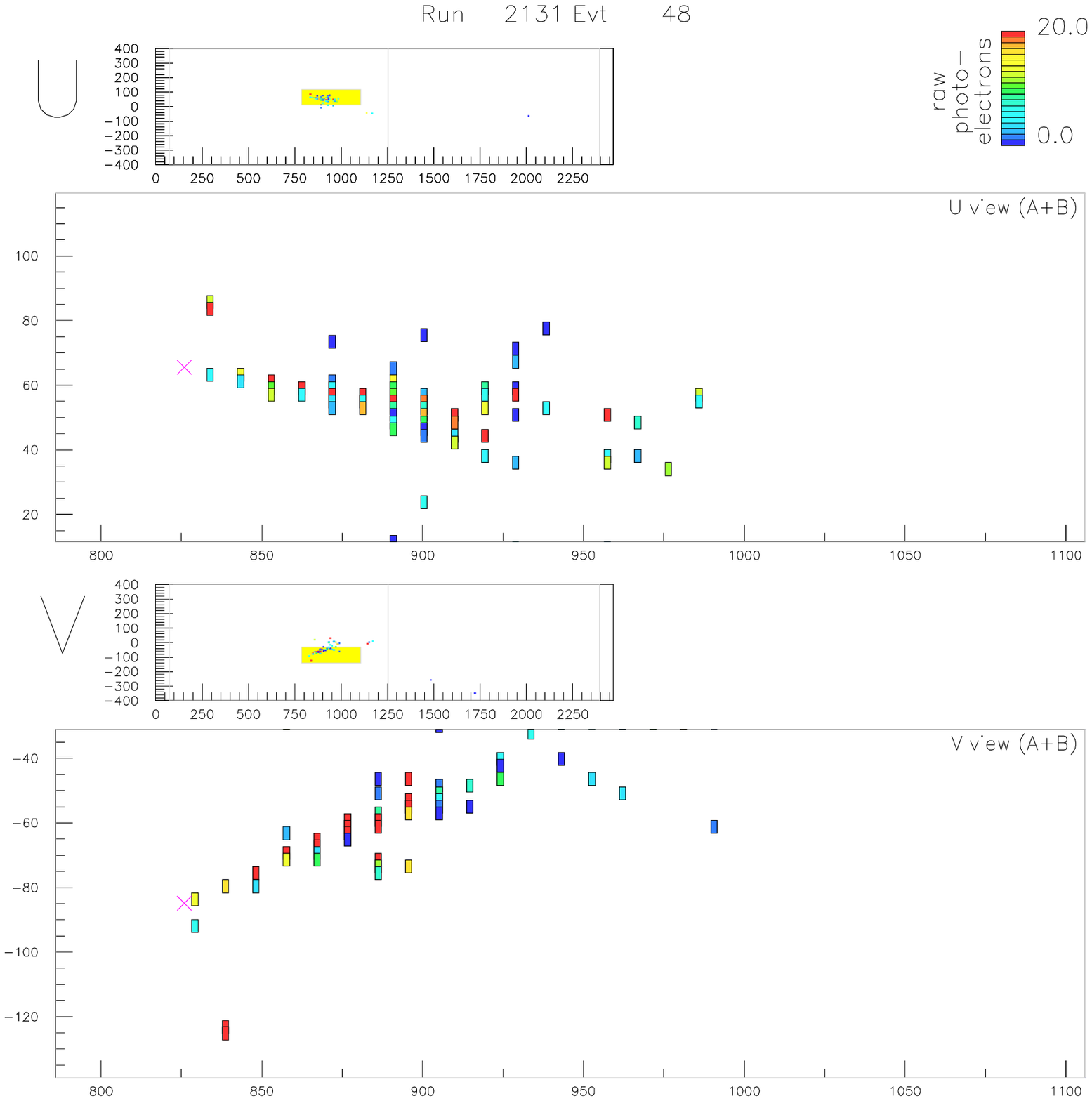,width=8.cm}
\epsfig{file=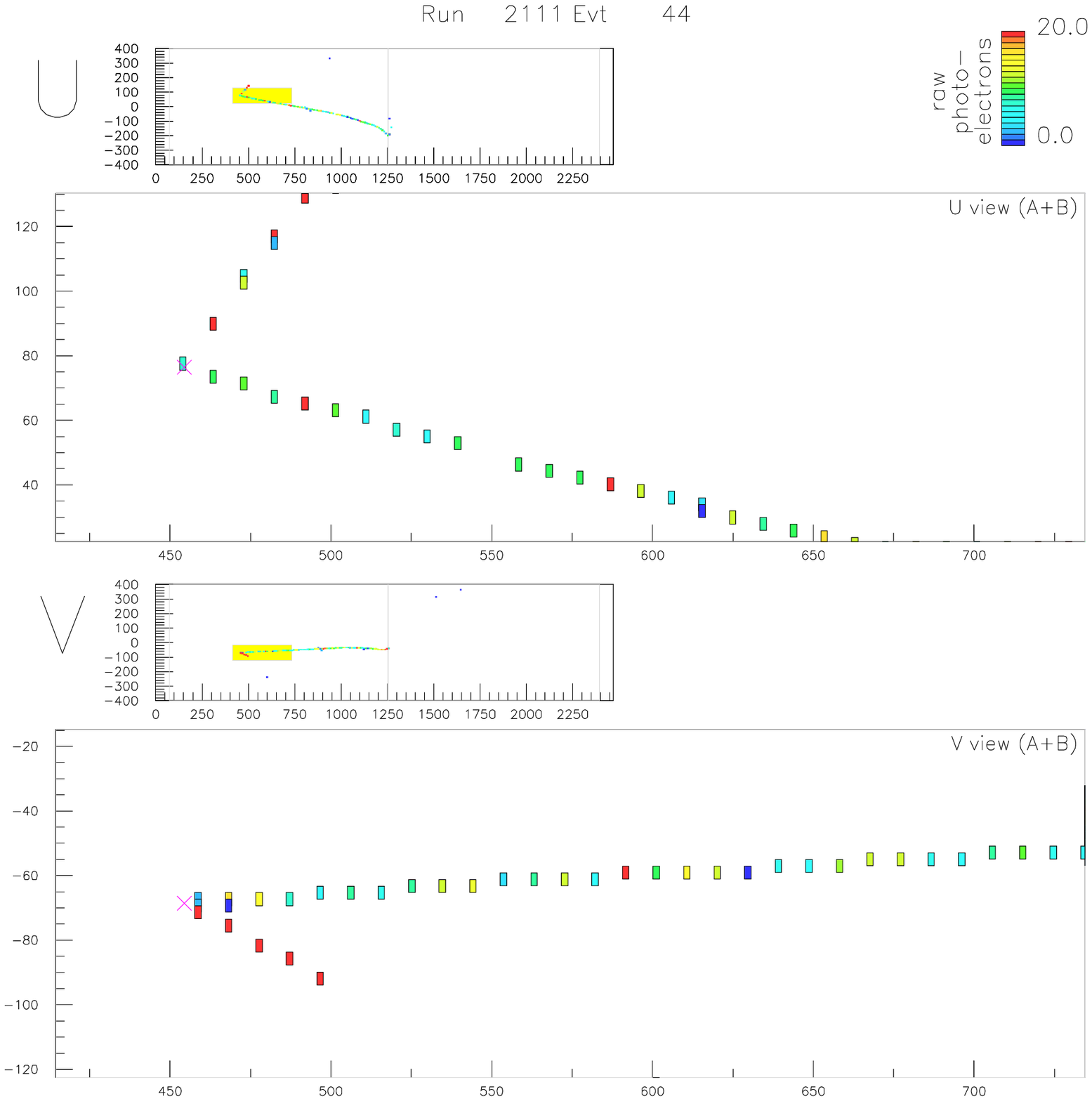,width=8.cm}
\epsfig{file=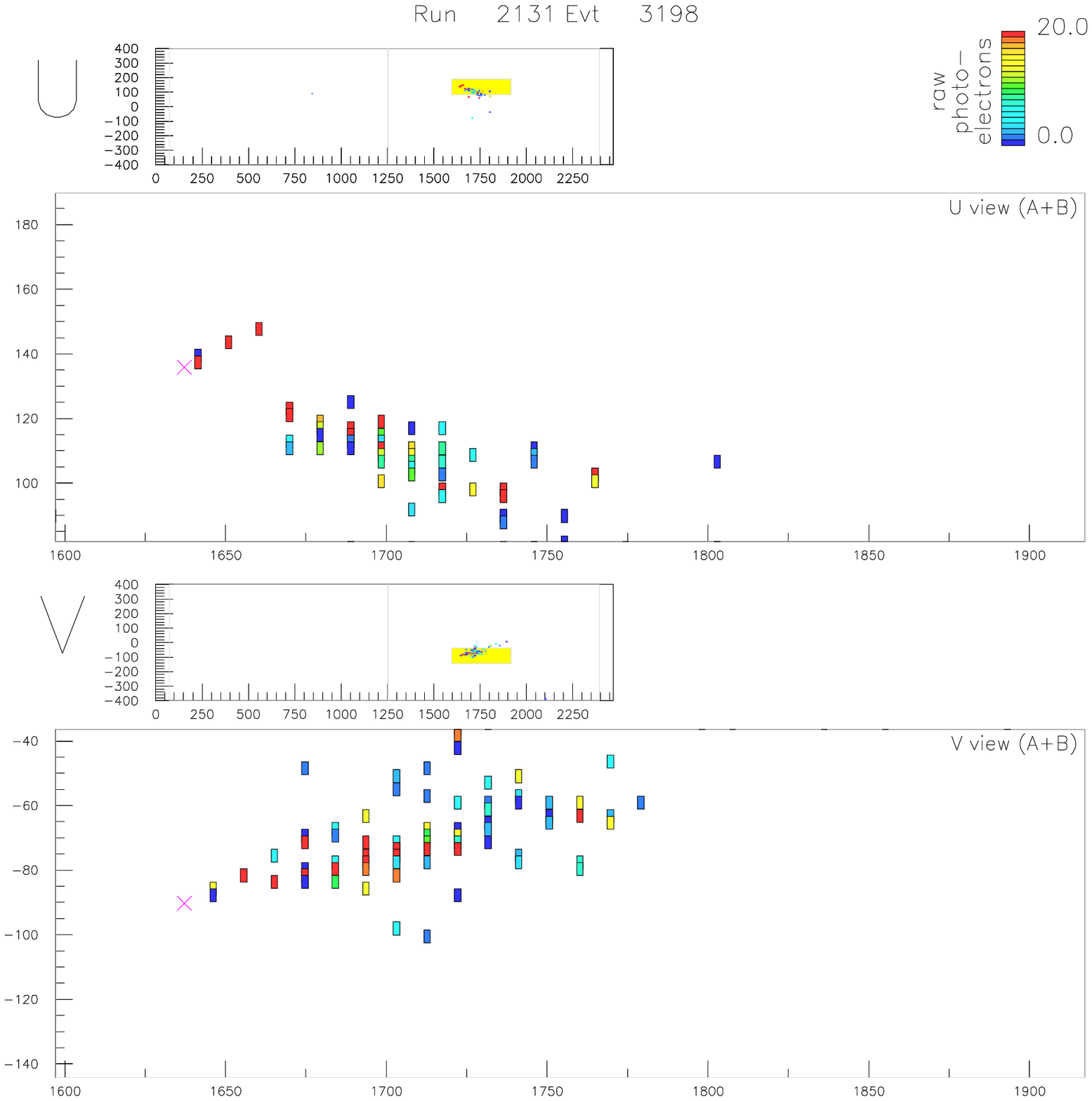,width=8.cm}
\end{center}
\caption{\label{display}
Typical events for $\nu_e$ CC,  $\nu_\mu$ CC, and NC.
}
\end{figure}

GMINOS produces an ADAMO-structured output file, allowing a data
analysis under the MAW (MINOS interface to PAW) package or a dedicated
event reconstruction program; it also offers event displays in the
longitudinal and face-on projections via mhpd.  In Fig.~\ref{display} 
we show some ``typical" events.

Running versions of GMINOS exist for both Linux and IRIX64 platforms.
All detector optimization and performance studies herewith referred to
were done by appropriate input card settings within the GMINOS
framework.

\subsubsection{Event Reconstruction}

In a highly segmented calorimetric detector, a 1-2~GeV electromagnetic
shower will leave hits in typically 10-20 consecutive planes, making
possible track finding in each view.  Muons of more than 0.5~GeV
produce at least 40 planes long tracks; identifiable tracks, although
usually shorter, are also often produced by charged pions and recoiling
protons.  High transverse segmentation provides good separation of two
close tracks; preliminary studies in which signatures of single 1-2~GeV
$\pi^0$'s were examined revealed that about 2/3 of them produce two
separable showers in at least one view.  This result is over a factor 2
better than obtained from the same study for the MINOS far detector.
On the other hand, a still finer segmentation, either transverse or
longitudinal, was found not leading to a significant performance
improvement.  In Table~\ref{seg} shows the  $\pi^0$ identification
capabilities for different segmentations.

\newpage

\begin{table}[ht]
\begin{center}
\begin{tabular}{|c|c|c|c|}
\hline
 Iron     & Scintillator & $\pi^0$ id & $e^-$ id   \\
thickness & strip width  & efficiency & efficiency \\ \hline
2.54 cm   &    4 cm      &   30\%     &   90\%     \\
1.00 cm   &    2 cm      &   50\%     &   90\%     \\
0.45 cm   &    3 cm      &   59\%     &   90\%     \\
0.45 cm   &    2 cm      &   66\%     &   90\%     \\
0.45 cm   &    1 cm      &   67\%     &   90\%     \\
0.23 cm   &    2 cm      &   69\%     &   90\%     \\
\hline
\end{tabular}
\caption{\label{seg}
$\pi^0$ identification capabilities for different segmentations.
}
\end{center}
\end{table}

  Most detector optimization studies, that
originally were done on a particular example of the steel and
plastic scintillator design, are easily extendible to alternative
designs, after defining what the equivalent segmentation is, expressed
in absorber radiation lengths (longitudinally) and Moli{\`e}re radii
(transversely).  In case of a high-Z absorber, gain in performance can
be obtained by including a few cm long separation (air gaps) between
each two adjacent planes.  This has the effect of increasing the 
effective mean radiation length of the detector and therefore improving
the angular resolution of tracks.
Dedicated simulations of the same 
basic steel/scintillator detector
design with different air gaps revealed a steady
increase of signal reconstruction efficiency  as the air  
gaps lengths increased from 0 to 3 cm,
followed by a situation where further improvement of angular resolution
merely compensates the losses due to increasingly poor clustering of
electromagnetic showers.

Event reconstruction is done entirely within the reconstruction
program included in the GMINOS package.  It consists of track fitting
and applying selection criteria at both the track and the event level.
Tracks are fitted and examined in each view separately.  A good track
is required to give hits in at least 4 planes and have good $\chi^2$
for a straight line.  Long muon tracks, which are often not straight
lines, are retrieved as being composed of two to several straight
segments.  Further analysis relies largely
on simple track characteristics like length, width and energy.

\subsubsection{$\nu_\mu$ CC and NC Identification}
\label{numu_dissaperance}

Events that fail the criteria imposed for $\nu_e$ appearance search,
to be described in the next section,  
are largely dominated by NC and $\nu_\mu$ CC.  The following simple
algorithm allows a highly efficient separation of the two classes
of events.

\begin{figure}[here]
\begin{center}
\epsfig{file=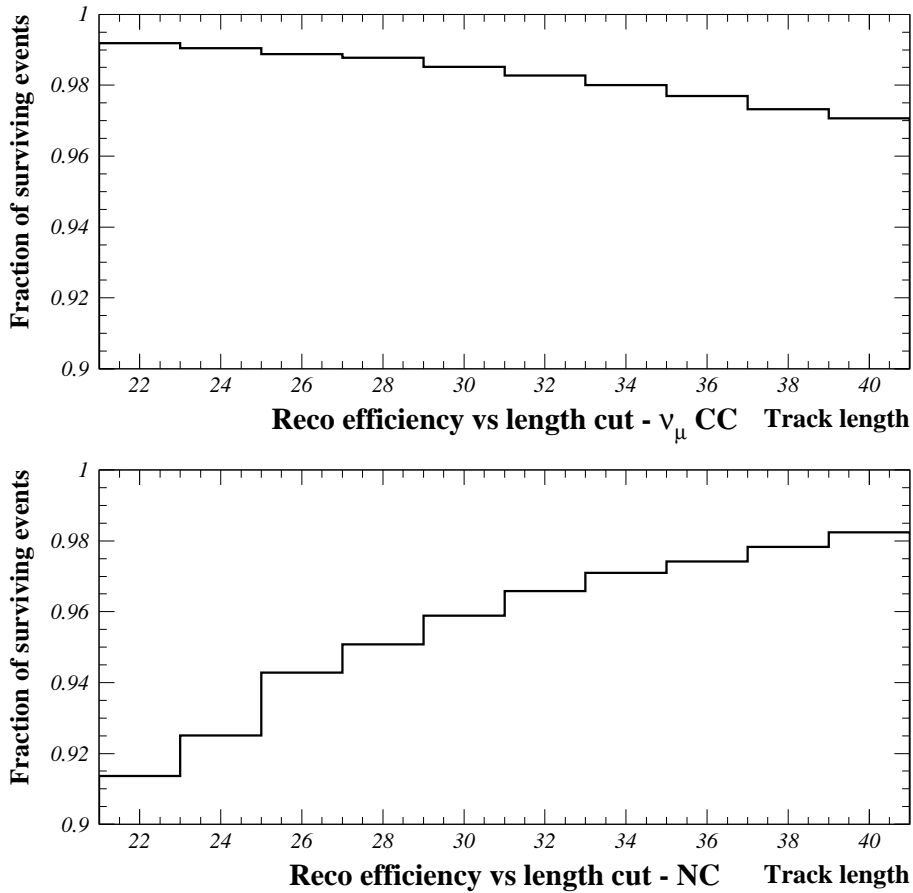,width=12cm}
\end{center}
\caption{\label{eff_cc_nc}
Efficiency for the identification of muons (top) and NC
(bottom), assuming low energy beam and a 10 km off-axis detector,
as a function of the track length cut value applied.}
\end{figure}

For muons of at least 0.5~GeV, the length of the produced track is
to a good approximation proportional to the initial muon energy.
A typical muon track is about 52 planes per view per GeV long and
essentially one cell wide.  In the NC events, only charged pions
can produce similar tracks, but with an incident beam spectrum
peaking at 2~GeV it is kinematically unlikely to have a charged
pion produce a 40 planes per view long track.  It is therefore
sufficient to concentrate on one cell thin tracks and check for
the longest track in the event to obtain two high purity samples,
dominated by $\nu_\mu$ CC and NC, respectively.  The identification
efficiency for $\nu_\mu$ CC and for NC, depending on the actual
value of the track length cut that is applied, is given in Fig.~\ref{eff_cc_nc}.
For example, applying a cut at 40 planes per view, one gets a 97\%
overall muon identification efficiency and $>$98\% efficiency for
NC.  The resulting samples for the medium  energy beam at baseline
of  900~km  and a 11.5~km off-axis detector are given in Table~\ref{info_rec}.

\begin{table}[htbp]
\begin{center}
\begin{tabular}{|c|c|c|c|}
\hline
Sample & $\nu_\mu$ CC & NC & $\nu_e$ CC \\ \hline
L$>$40 &  34.1        & 0.7  &  0. \\
L$<$40 &   1.1        & 39.3 &  2.6 \\ 
       &              &      &      \\ \hline
\end{tabular}
~~~~~~~~~~~~~~~~\begin{tabular}{|c|c|c|c|c|}
\hline
       & $\nu_e^{signal}$ & $\nu_e^{beam}$ & NC & $\nu_\mu$ CC \\ \hline
All    &  1.28      &  1.04  &  24.95  & 18.04  \\
$\epsilon$ &  0.43 &  0.10 &  0.005 &  0.0004 \\
Final  &  0.55     &  0.10  &  0.11   & 0.007   \\ \hline
\end{tabular}
\caption{\label{info_rec} Reconstruction efficiencies. LEFT:  $\nu_\mu$ CC
and NC event reconstruction as a function of the track length, L.
RIGHT: for $\nu_e$  CC event reconstruction  for energies between 1.5 to 3~GeV.
We  gives expected number of events for 1~kTon-year for a detector located 
at 11.5~km from the axis, and at a length of 900~km.  $\nu_\tau$ CC events 
do not contribute because the beam energy is below the threshold of that 
reaction.
}
\end{center}
\end{table}

\begin{figure}[here]
\begin{center}
\epsfig{file=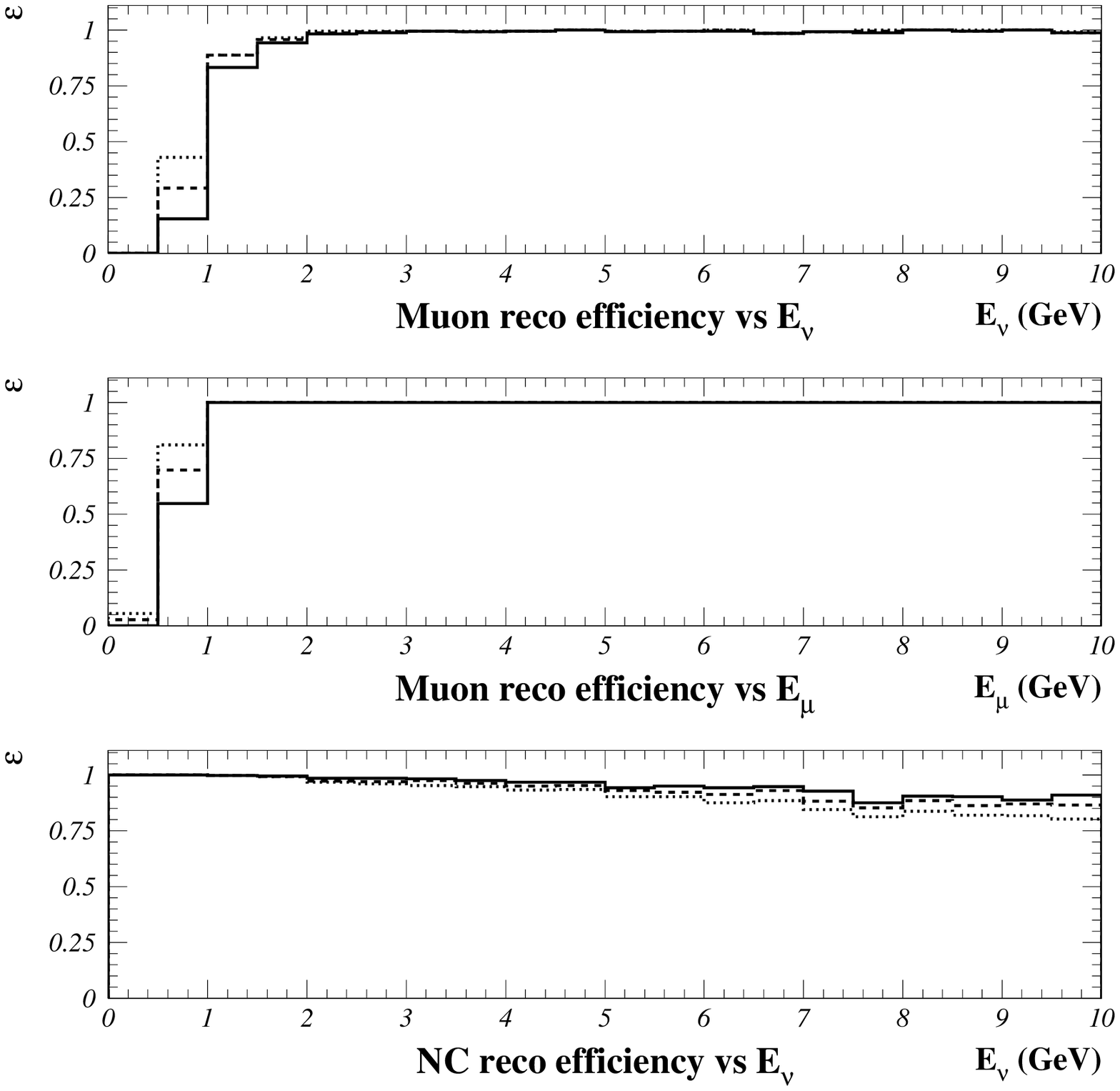,width=10cm}
\end{center}
\caption{\label{survive_cc_nc} 
Muon identification efficiency as a function of beam
energy (top), as a function of muon energy (middle), and NC
identification efficiency as a function of beam energy (bottom).
Solid lines correspond to a track length cut at 40 planes per view,
dashed lines to 36 planes, dotted to 32 planes.}
\end{figure}

As is found, background in the $\nu_\mu$ CC sample amounts to
merely 2\%.  The background contamination of the NC sample
depends significantly on the amount of $\nu_e$ appearance and,
as usual, we have arbitrarily assumed $|U_{e3}|^2=0.01$.
Fig.~\ref{survive_cc_nc} shows the muon efficiency as a function of the incident
neutrino energy (top), as a function of muon energy (middle), and
the NC efficiency as a function of neutrino energy (bottom), for
three values of the track length cut.

\begin{figure}[here]
\begin{center}
\epsfig{file=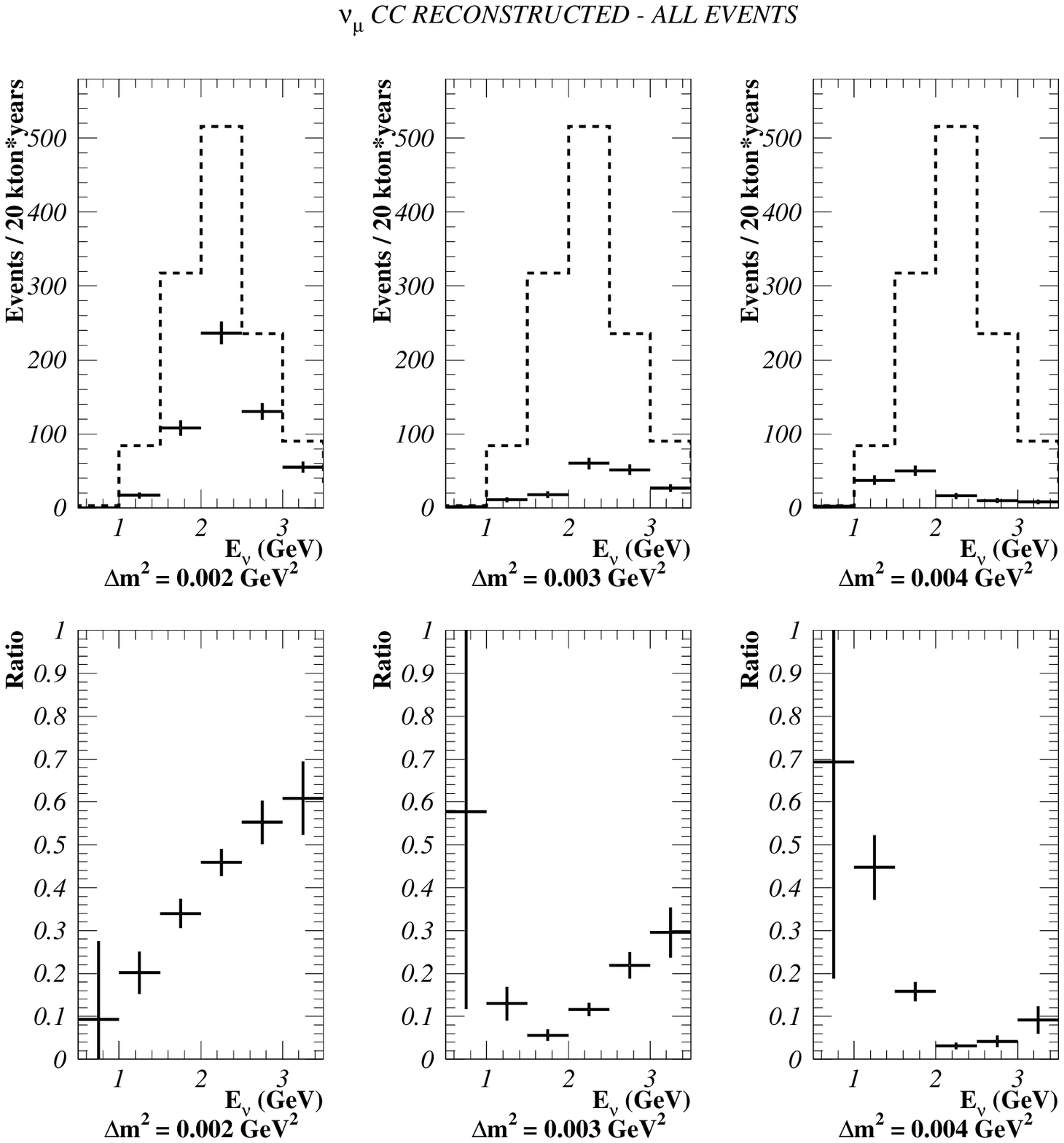,width=10cm}
\end{center}
\caption{\label{cc_all}
Reconstructed energy distributions for all $\nu_\mu$ CC
events for three different values of $\Delta m^2_{23}$.  Also shown is
the measured spectrum for the no oscillation case (dashed histograms)
and respective ratios (lower plots).}
\end{figure}

\begin{figure}[here]
\begin{center}
\epsfig{file=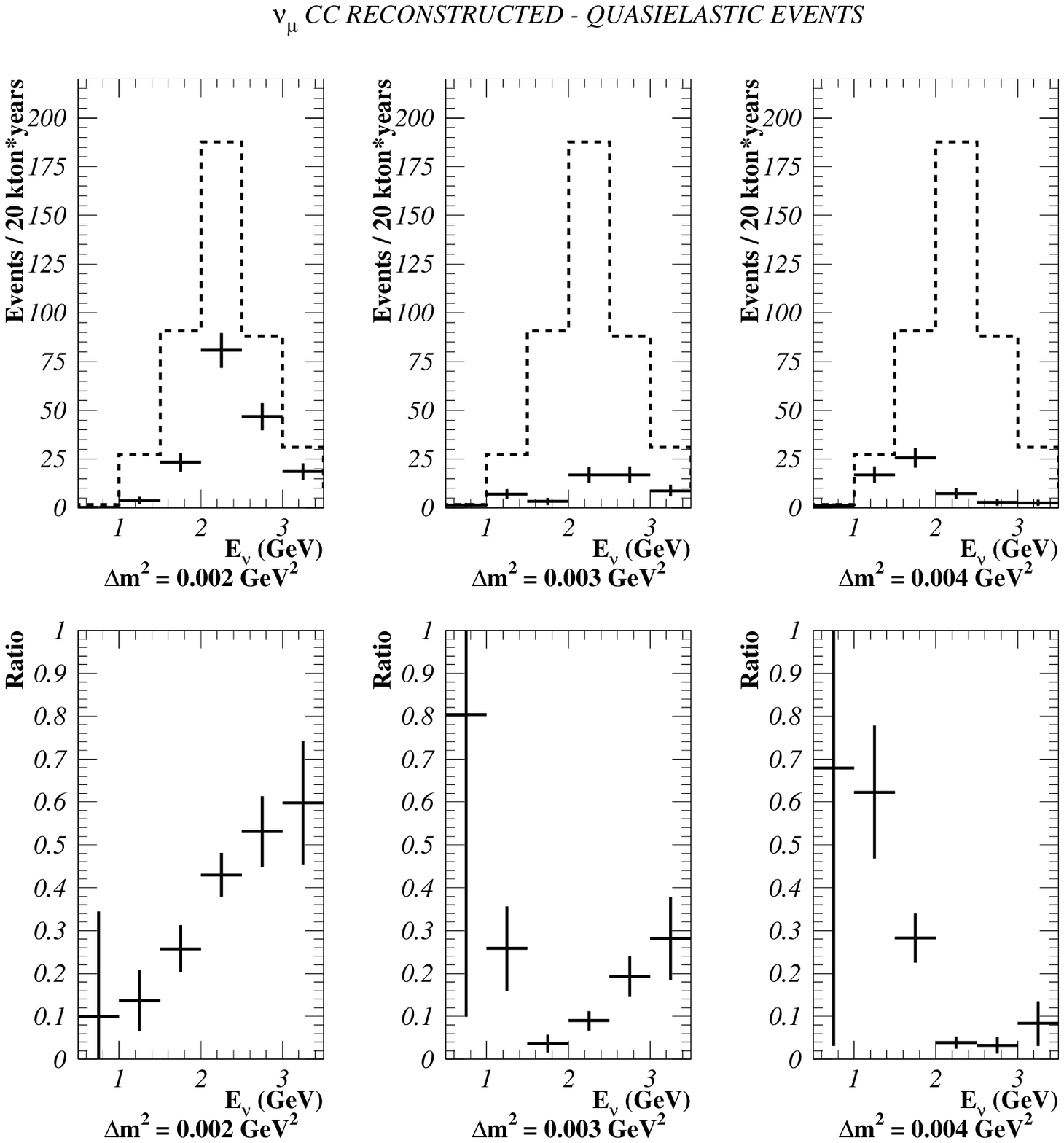,width=10cm}
\end{center}
\caption{\label{cc_qe}
Reconstructed energy distributions for quasi-elastic $\nu_\mu$ CC
events for three different values of $\Delta m^2_{23}$.  Also shown is
the measured spectrum for the no oscillation case (dashed histograms)
and respective ratios (lower plots).}
\end{figure}

From the track length, the muon energy can be determined to better
than 10\% for fully contained events.  This corresponds to a similar
accuracy in the beam energy measurement based on quasi-elastic events
(and those inelastic events in which final state hadrons are not
energetic enough to leave identifiable tracks).  Extending the
analysis to all events improves the statistical power by a factor
2-3, but the accuracy of neutrino energy reconstruction is limited
by the uncertainty in energy measurement of hadronic and
electromagnetic showers and is of $\sim$20\%.  Whether one or the
other approach will be preferred depends mainly on the systematic
uncertainties related to smearing and selection efficiency corrections
that will require more detailed Monte Carlo studies, and to our
knowledge of the corresponding cross sections at the time of running
the experiment.  Fig.~\ref{cc_all}  shows
the reconstructed energy spectra of all $\nu_\mu$ CC for three values
of $\Delta m^2_{23}$ and Fig.~\ref{cc_qe}. the same spectra for quasi-elastic-like
events only.  Error bars are the statistical errors corresponding
to a 20 kTon$\times$year exposure (fiducial).  Clearly, from a purely
statistical
point of view a precise measurement of
$\Delta m^2_{23}$ is possible.

\subsubsection{$\nu_e$ CC Identification}
\label{nue_aperance}

Analysis towards $\nu_e$ appearance should be optimized in means of
the expected ``figure-of-merit", defined as $S/{\sqrt{S+B}}$, where
$S$ is the number of expected signal events under the (arbitrary)
assumption of $|U_{e3}|^2 = 0.01$ and $B$ the expected residual
background in the final sample per 1 kTon$\times$year.

Most tracks coming from charged pions and
recoiling protons can be rejected by requiring a mean track width
of at least two cells and a width at maximum of at least three cells.
``Baby tracks", that is, tracks found in the vicinity of the end of
a longer main track and less than half of its length, come mostly
from secondary particles within the same shower and are discarded.
Conversely, two tracks of comparable length and/or pointing at the
same interaction vertex are a signature of a NC event with a $\pi^0$
in the final state.

\begin{figure}[here]
\begin{center}
\epsfig{file=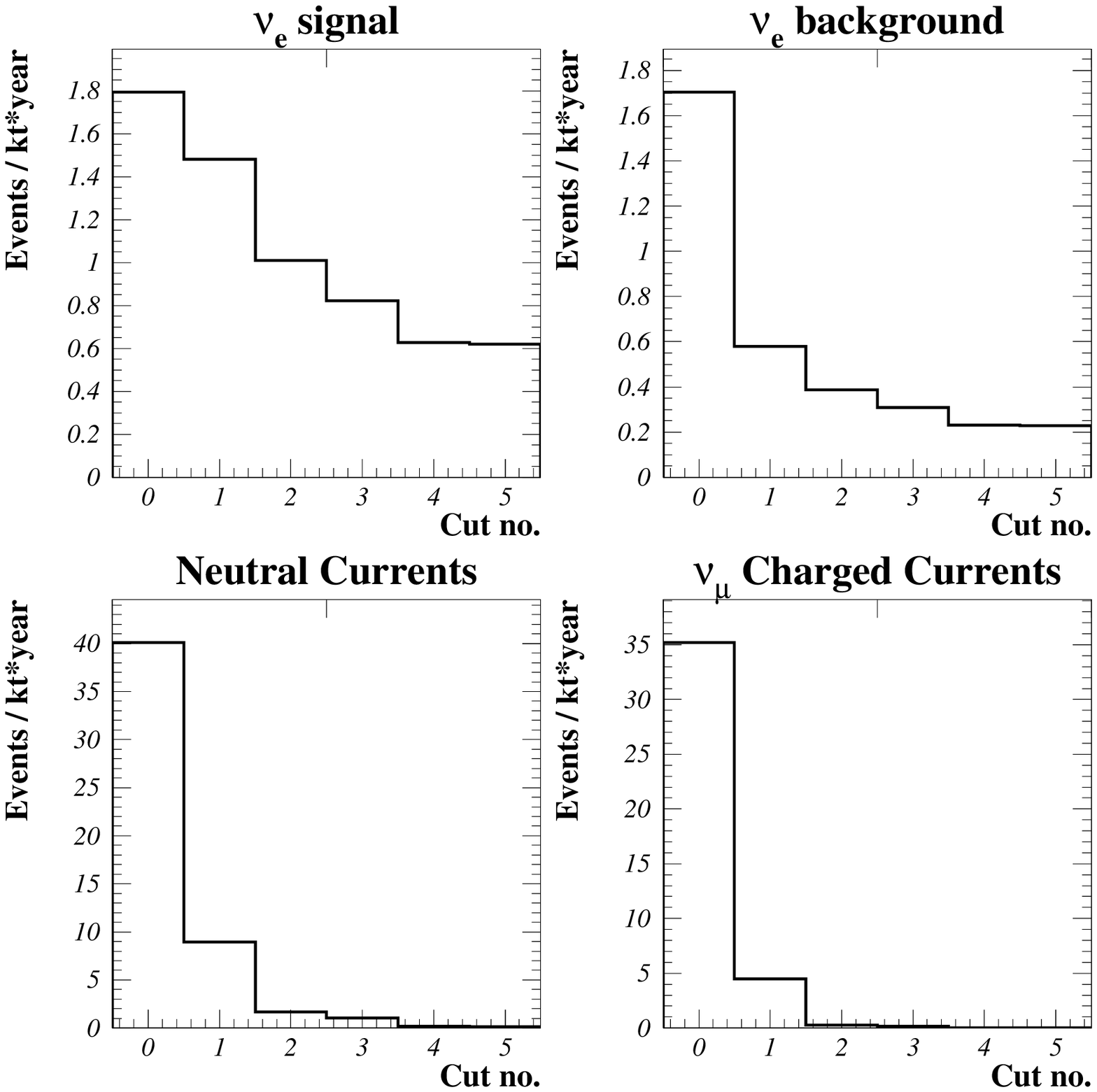,width=10cm}
\end{center}
\caption{\label{effy_nue}
Initial samples per 1 kTon$\times$year for four classes of
events: signal $\nu_e$ CC, background $\nu_e$ CC, NC and $\nu_\mu$ CC,
and their reduction rates after applying subsequent selection criteria.
The cuts are: 1. Visible energy, 2. One good track per view, 3. Small
$p_T$, 4. Large energy fraction carried by track, 5. Shower starting
from interaction vertex.}
\end{figure}

After track selection, a signal candidate event is required to leave
exactly one good track in each view.  Additional selection criteria
are imposed on the event basis.  To optimize the ratio of $\nu_e$
signal to intrinsic $\nu_e$ background, a window in the total visible
energy is defined (for $\Delta m^2_{23} = 0.0025 - 0.003$~GeV$^2$ a
reasonable choice is 1-3~GeV).  A small missing $p_T$ with respect to
the beam direction is required, a minimum fraction of the total event
energy carried by the track (both criteria helping reject NC), and
no track longer than 28 planes in any view (suggestive of a muon).
Remaining NC background is further reduced by checking for a
displacement of the beginning of the shower with respect to the
interaction vertex, the latter being identified by the trace of
the recoiling proton (if any).  The efficiency of the above criteria
on generated samples of signal $\nu_e$ CC, background $\nu_e$ CC,
NC and $\nu_\mu$ CC, is depicted in Fig.~\ref{effy_nue}.

\begin{figure}[here]
\begin{center}
\epsfig{file=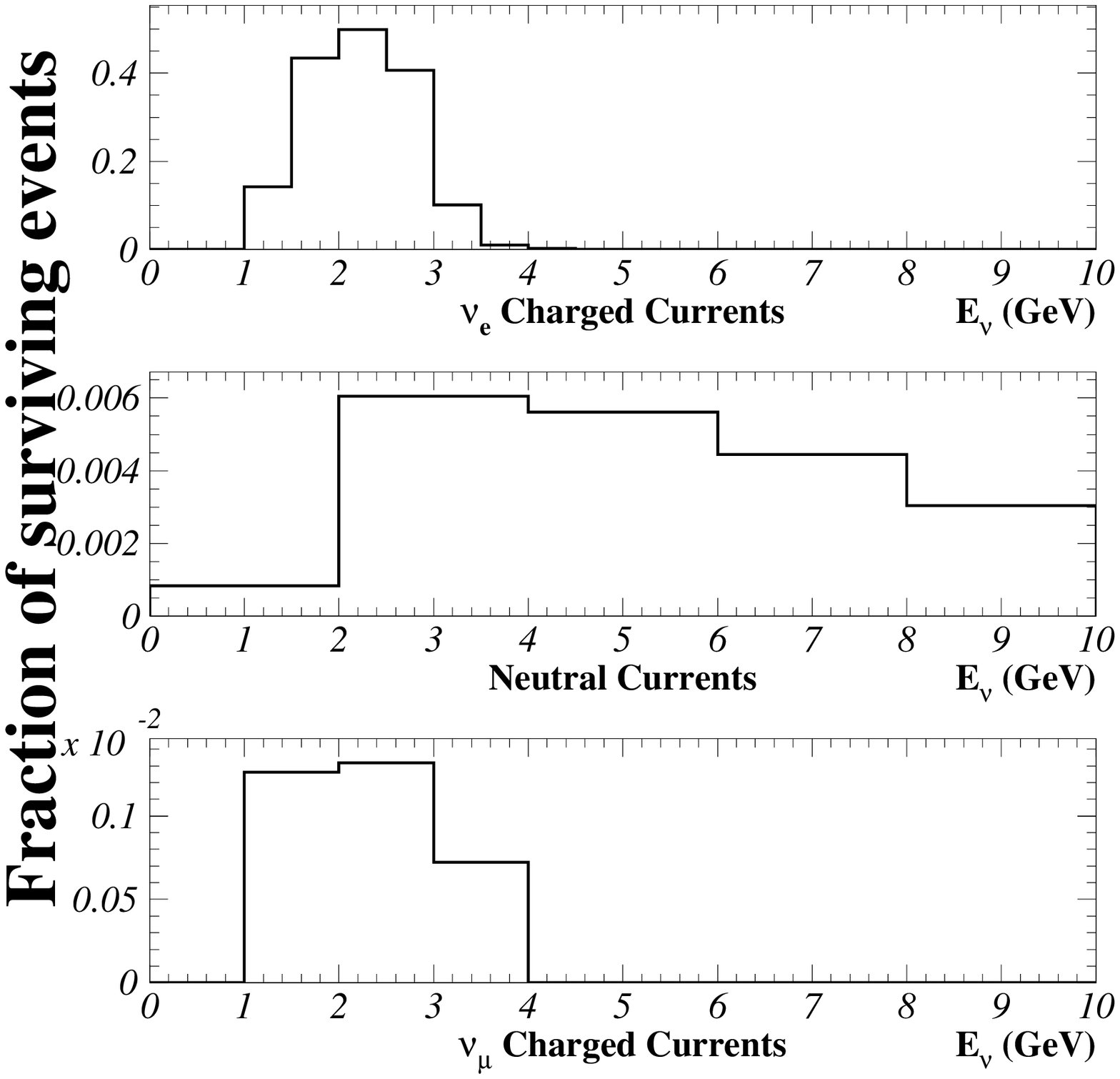,width=10cm}
\end{center}
\caption{\label{survive_nue}
Fraction of surviving events for the four classes of events,
as a function of beam energy.}
\end{figure}

The resulting reconstruction efficiency and remaining contaminations
from NC and $\nu_{\mu}$ CC as a function of the incoming neutrino energy
are shown in Fig.~\ref{survive_nue}.  To an approximation in which exactly the 
same analysis procedures and selection criteria are applicable,
these efficiencies should be convoluted with the actual beam spectrum
to obtain the overall reconstruction efficiencies and background
rejection levels for any assumed beam.
As is learned, for a typical NuMI off-axis beam under consideration,
one gets a total signal reconstruction efficiencies of 35-42\%,
a rejection of about 99.7\% of all NC events is
obtained, while $\nu_{\mu}$ CC are suppressed to a negligible level.
The total background is therefore dominated by the
intrinsic $\nu_e$ component of the beam and any further background
reduction is impractical.  As an example,
the total expected signal and background
rates for a 10 km off-axis detector in the low energy NuMI beam
per 1 kTon$\times$year are given in table.

\begin{table}[ht]
\begin{center}
\end{center}
\end{table}

\subsection{Physics Sensitivity with $4\times 10^{20}$ Protons per year}
\label{Physics_Sensitivity1}

To fully appreciate the capabilities of a Proton Driver upgrade, it is 
first useful to consider what physics would be available with a 20\,kTon 
fine-grained calorimeter and a five year run, assuming $4\times 10^{20}$ 
protons on target per year, or the nominal NuMI intensity.  The plots 
that follow are for a beam that is 900~km from Fermilab, and located 
11.5~km perpendicular from the center of the beam.  

\subsubsection{$\nu_\mu$ Disappearance Measurements} 

First of all, 
in combination with the MINOS experiment, an off-axis 
experiment would dramatically improve the precision on the 
measurements 
of the atmospheric parameters, $\Delta m_{23}^2$ and $\theta_{23}^2$.  
As we will see in later sections, 
this level of precision is needed in order to be able to interpret
the $\nu_{e}$ appearance data properly, and if one is to ever extract 
information about the CP-violating phase of the mixing matrix.  

\begin{figure}[here]
\begin{center}
\epsfig{file=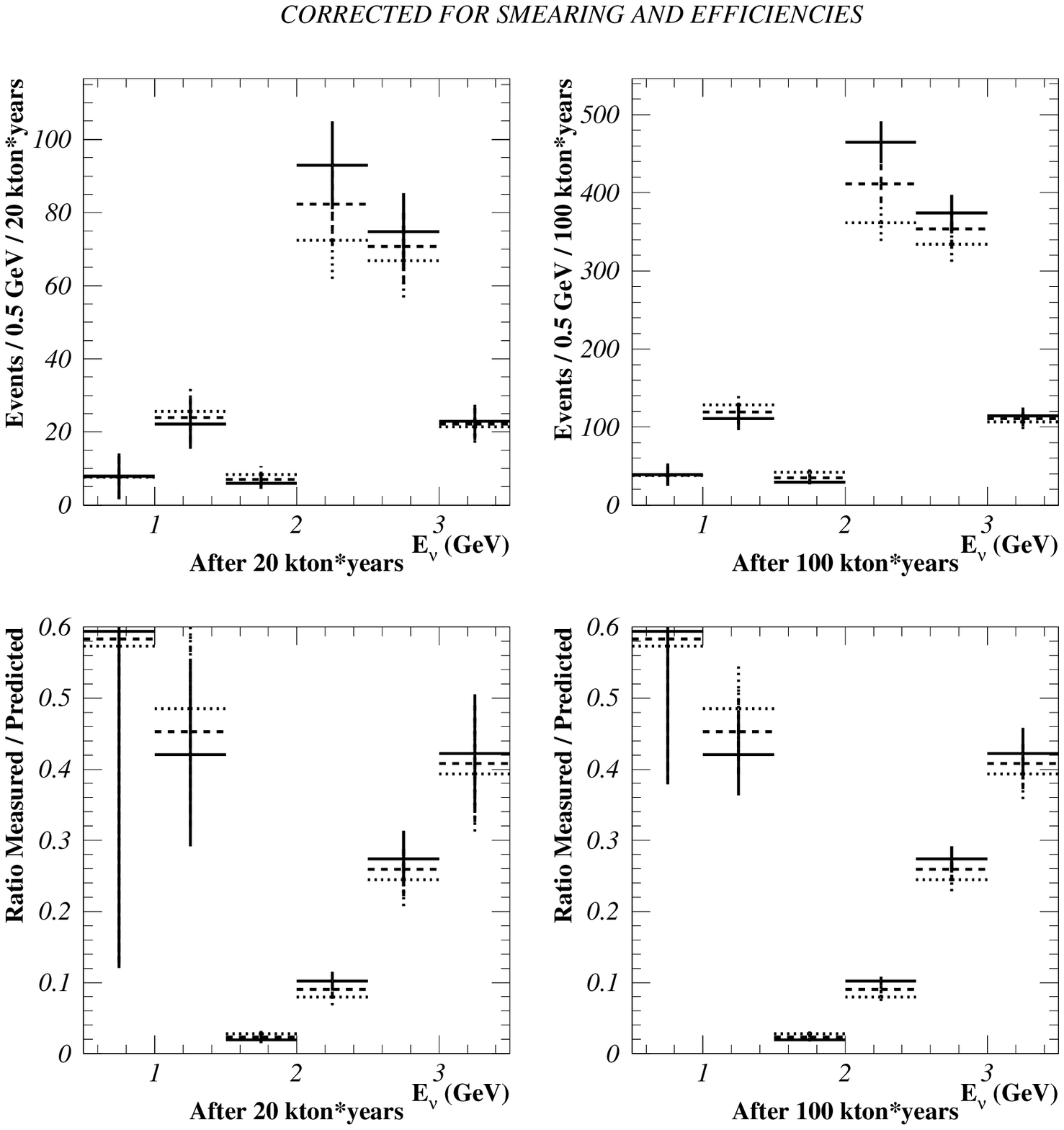,width=10cm}
\end{center}
\caption{\label{1percent}
Measured $\nu_\mu$ energy distributions
events for $\Delta m^2_{23}$=0.003 (solid lines), 0.00305 (dashed) and
0.0031 (dotted)~GeV$^2$ after an exposure of 20\,kTon$\times$years
and 100\,kTon$\times$years.  Also shown are the respective ratios
\mbox{measured/predicted}.  Only statistical errors are considered.}
\end{figure}

From the analysis described in section~\ref{numu_dissaperance}, we expect
that for a baseline of 900km and 
$\Delta m^2_{23}$=0.002\,GeV$^2$, the rate of $\nu_\mu$ disappearance 
is
close to 50\% and increases to 90\% if $\Delta m^2_{23}$=0.003\,GeV$^2$.
In the latter case, most disfavorable from the statistical point of
view, a 2\% measurement of $\Delta m^2_{23}$ is possible after an exposure
of 20\,kTon$\times$years, assuming that systematics will also be under
control to an appropriate level.  Figure~\ref{1percent}  depicts the measured
$\nu_\mu$ spectra for 3 close values of $\Delta m^2_{23}$ near 0.003~GeV$^2$
and for two different exposures,
after necessary corrections for energy smearing and selection
efficiency, and the corresponding ratios between measured and predicted values.
A simple fit to the latter yields a value of $\Delta m^2_{23}$ with
a statistical error of $\pm$0.0042~GeV$^2$ for a 20\,kTon$\times$year
exposure and of $\pm$0.0019~GeV$^2$ for a 100\,kTon$\times$year exposure.
This infers a measurement to better than 2\% and 1\%, respectively,
with purely statistical uncertainties being considered.

\subsubsection{$\nu_{e}$ Appearance Search}

\begin{figure}[htp]
\centerline{\epsfxsize 13.2cm \epsffile{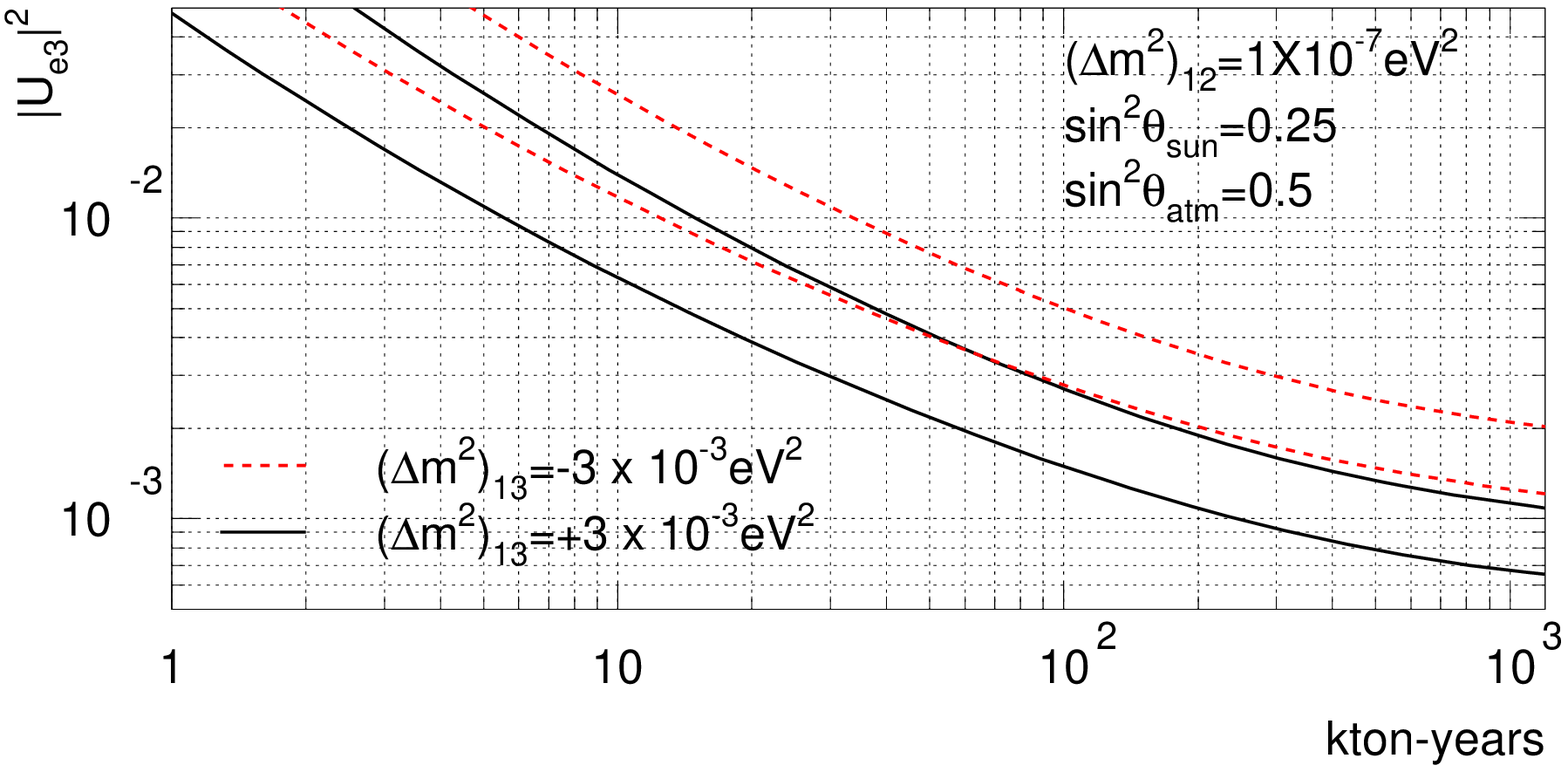}}
\caption{Two and three sigma sensitivity reach for $|U_{e3}|^2$ as a function of the
running time (in kTon-years), for a normal neutrino hierarchy (solid lines) and an inverted
neutrino hierarchy (dashed lines). $|\Delta m^2_{23}|= 3\times 
10^{-3}$~eV$^2$, $\sin^2\theta_{\rm atm}=1/2$ and $\Delta m^2_{12}=
10^{-7}$~eV$^2$, $\sin^2\theta_{\odot}=1/4$, $\delta=0$.
For a normal neutrino mass hierarchy, a 20~kTon detector with (without) a
Proton Driver upgrade will be able to exclude at a  two sigma level 
$\nu_\mu\to\nu_e$ oscillations, if
$|U_{e3}|^2\le 0.00085(0.0015)$ after 5 years.}
\label{kton-year}
\end{figure}

For the experimental parameters listed above, an off-axis experiment 
with a fine-grained calorimeter could expect to improve the limits on 
$\nu_\mu \to \nu_e$ by roughly an order of magnitude past what has 
already been set by the CHOOZ reactor experiment.  Although 
Fig.~\ref{detector1} showed the limits one could achieve in a simple model 
with no matter effects and several detector options, Fig.~\ref{kton-year} 
shows for the fine-grained calorimeter described above
what the same plot looks like neglecting the solar mass splitting contribution,
but including both possibilities of matter effects.

First, one should determine the sensitivity to observing $\nu_e$-appearance
as a function of time (in kTon-years of detector exposure to the beam).
Observing a signal depends not only on  the value of $|U_{e3}|^2$ but also
on the neutrino mass hierarchy.
Fig.~\ref{kton-year} depicts the two and three sigma
sensitivity to $|U_{e3}|^2$ (see~\cite{offaxis} for details) as a
function of the number of kTon-years of accumulated neutrino beam   
data collected off-axis, in the case $\Delta m^2_{23}=\pm 3\times 
10^{-3}$~eV$^2$, $\sin^2\theta_{\rm atm}=1/2$ and $\Delta m^2_{12}=
10^{-7}$~eV$^2$, $\sin^2\theta_{\odot}=1/4$.
First, in order to be sensitive to values of
$|U_{e3}|^2$ which are significantly smaller than the current 
CHOOZ bound ($|U_{e3}|^2\lesssim 0.05$~\cite{CHOOZ}), one is required to 
accumulate more than 40~kTon-years of data, in the case of a normal hierarchy,
or more than 150~kTon-years in the case of an inverted hierarchy.
This means, assuming the nominal NuMI beam, roughly two or eight 
years of running with a 20~kTon detector.  
Note that the $10\%$ uncertainty 
on the background determination dictates that, even after accumulating
an infinite amount of statistics, the three sigma 
reach of the off axis experiment plateaus at around $|U_{e3}|^2\sim 1\times 
10^{-3}$ ($2\times 10^{-3}$) for a normal (inverted) hierarchy.
As is clear from Fig.~\ref{kton-year}, 
in the case of an inverted hierarchy, the sensitivity is
significantly worse. This is also expected, since matter effects enhance
the $\nu_e$ appearance rate in the case of a normal hierarchy and
reduce it in the case of an inverted hierarchy.

Note that the sensitivity would be significantly different 
for different values of $|\Delta m^2_{23}|$ and that, by design, the 
sensitivity is optimal at around $|\Delta m^2_{23}|\sim 3\times 
10^{-3}$~eV$^2$. We have verified that it 
does not deteriorate significantly for 
$|\Delta m^2_{23}|\sim (2-4)\times 10^{-3}$~eV$^2$.

\subsection{Physics Sensitivity with $16\times 10^{20}$ Protons per year}

If there is a Proton Driver upgrade with as much as a factor of four 
improvement above the ``nominal'' proton rate, then the physics possibilities
become far richer.  However, in order to accurately describe just how much 
physics one can do, it becomes necessary to specify first what the solar 
mass splitting is.  Simply put, if the solar mass splitting is below 
$10^{-5}$, then an experiment 
can still further the search for $\nu_\mu \to \nu_e$, 
and may even have 
the sensitivity to determine the neutrino mass hierarchy.  If 
the solar mass splitting is higher, then signals become far more complicated,
but these complications are due to the CP-violating terms, and as such are 
more than welcome!  In this section we will outline the physics reach 
for three different scenarios:  1: where the solar mass splitting is 
too small to be measured by KamLAND, 2: where the solar mass splitting is 
measured at the 10\% level by KamLAND, and finally 3:  when the solar mass
splitting is so large that KamLAND cannot accurately measure it, although it 
would see firm evidence for $\nu_e$ disappearance.

If the solar mass splitting is significantly below
$10^{-4}$, then a Proton Driver upgrade would provide another factor of 
two in reach for $|U_{e3}|^2$, if $\nu_\mu \to \nu_e$ 
has not already been seen, as shown in Fig.~\ref{measure_low}.  

\subsubsection{$\Delta m^2_{12}\ll 10^{-5}$~eV$^2$}

If KamLAND does not observe a suppression of the reactor anti-neutrino 
flux, the LMA solution to the solar neutrino puzzle will be excluded 
\cite{Kam,Kam_sim}, indicating that $\Delta m^2_{12}\ll 10^{-5}$~eV$^2$ and/or 
$\tan^2\theta_{\odot}\ll 1$. In this case, it is well known that
the CP-odd phase $\delta$ is not observable in standard long-baseline
experiments, not only because solar oscillation do not have enough time
to ``turn on,'' but also because matter effects effectively prohibit
any neutrino transition governed by the solar mass-squared difference.
This being the case, one can only study 
$\nu_{\alpha}\leftrightarrow\nu_{\beta}$ transitions governed by 
the atmospheric mass-squared difference. 

If $\nu_\mu \to \nu_e$ has not been observed, then the additional 
protons will be crucial to improve the search for a non-zero $|U_{e3}|^2$,
as shown above.  However, if $\nu_\mu \to \nu_e$ has been observed at 
at least the three sigma level, then a Proton Driver upgrade would allow
one to get statistics in anti-neutrino running in a relatively modest running
time, and determine the neutrino mass hierarchy.  

\begin{figure}[htp]
\centerline{\epsfxsize 13cm \epsffile{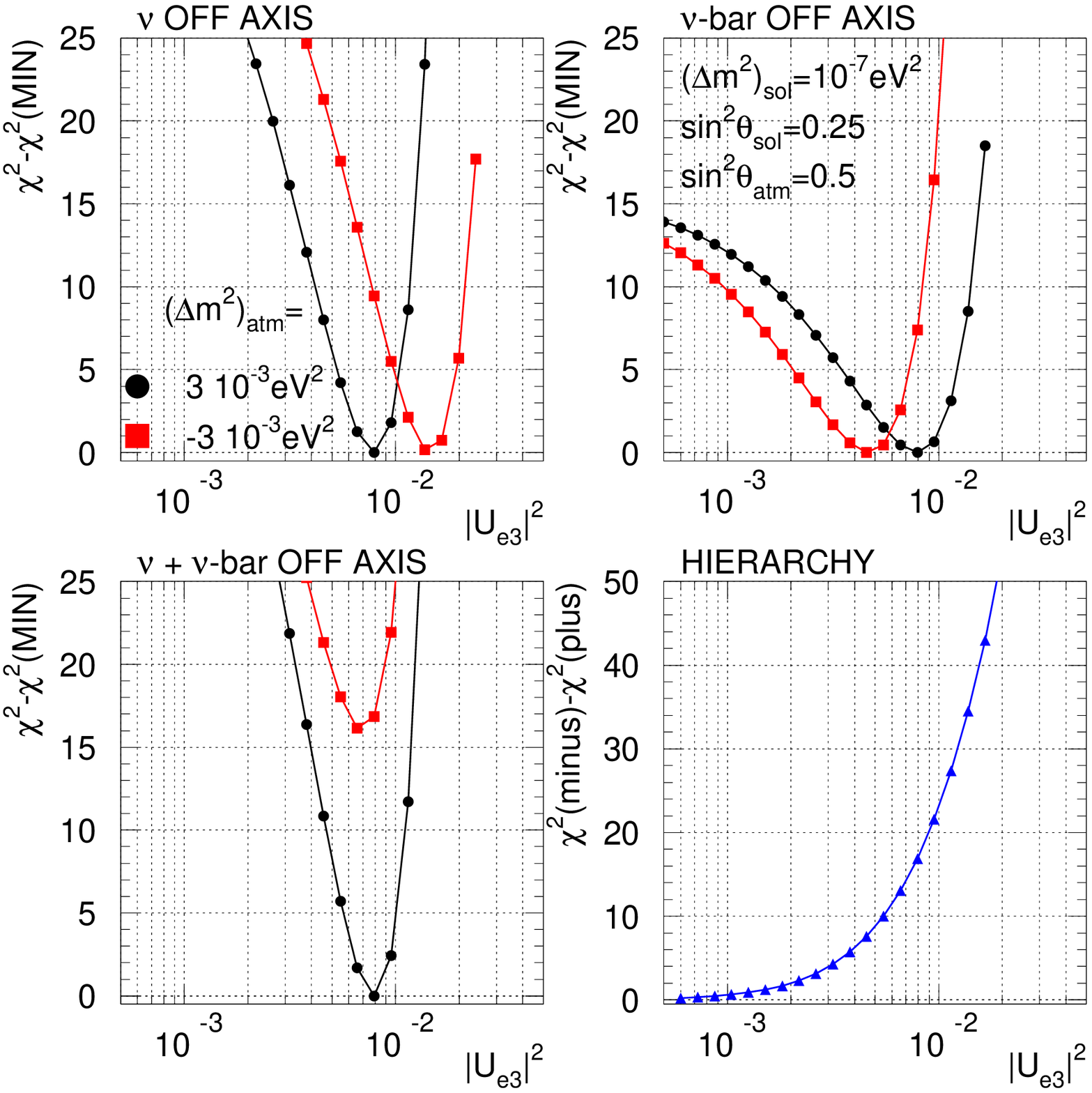}}
\caption{top-($\chi^2-\chi^2_{\rm min}$) as a function of $|U_{e3}|^2$, assuming both neutrino
mass hierarchies, upon analyzing simulated data consistent with $\Delta m^2_{23}>0$ and
$|U_{e3}|^2=0.008$ after 120\,kTon-years of neutrino-beam running (left) and 
300\,kTon-years
of anti-neutrino-beam running (right). bottom,left-same as above, after combining the
two data sets. bottom,left-difference of minimum value of $\chi^2$ obtained with the hypothesis
$\Delta m^2_{23}>0$ and $\Delta m^2_{23}<0$ as a function of the input $|U_{e3}|^2$. See text
for details. $|\Delta m^2_{23}|= 3\times 
10^{-3}$~eV$^2$, $\sin^2\theta_{\rm atm}=1/2$ and $\Delta m^2_{12}=
10^{-7}$~eV$^2$, $\sin^2\theta_{\odot}=1/4$, $\delta=0$.}
\label{measure_low}
\end{figure}

Consider what happens if 
one detects an excess of $\nu_e$-like events: the next step in principle
would be to 
determine the value of $|U_{e3}|^2$. One can do this by performing
a $\chi^2$ fit to the ``data''.  It is assumed that the atmospheric
parameters $|\Delta m^2_{23}|=3\times 10^{-3}$~eV$^2$, 
$\sin^2\theta_{\rm atm}=1/2$
are precisely known.  Fig.~\ref{measure_low}(top,right)
depicts $\chi^2-\chi^2_{\rm MIN}$ as a function of $|U_{e3}|^2$ 
corresponding to 120~kTon-years\footnote{This corresponds to
six years of running with the current NuMI beam configuration and
a 20\,kTon detector. With a Proton Driver, however, the same amount of
data can be collected in 1.5 years. This will become crucial later.} 
of ``data'' collected with 
a neutrino beam (as defined earlier, the neutrino (anti-neutrino) 
beam consists predominantly of 
$\nu_{\mu}$ ($\bar{\nu}_{\mu}$)). Note that, while the data were simulated with
$\Delta m^2_{23}=+3\times 10^{-3}$~eV$^2$ and $|U_{e3}|^2=0.008$, a different
solution, with the same goodness of fit, is found for $\Delta m^2_{23}
=-3\times 10^{-3}$~eV$^2$, $|U_{e3}|^2=0.015$.\footnote{It is important
to reemphasize that $|\Delta m^2_{23}|$
is {\sl not} a fit parameter. It is assumed to be known from different
sources, such as the study of the $\nu_{\mu}\rightarrow\nu_{\mu}$
disappearance channel in the off-axis experiment, discussed in the previous subsection.} 
This implies that
if the neutrino mass hierarchy is not known, instead of obtaining a precise
measurement of $|U_{e3}|^2=0.008\pm 0.0025$ (these are two sigma error bars), 
one is forced to quote a less precise (very non-Gaussian) 
measurement: $0.0055<|U_{e3}|^2<0.018$ at the 
two-sigma confidence level.   

Fig.~\ref{measure_low}(top,right) depicts $\chi^2-\chi^2_{\rm MIN}$ 
as a function of $|U_{e3}|^2$ corresponding to 300\,kTon-years of 
``data'' collected with the anti-neutrino beam, which, assuming a 20\,kTon 
detector and a Proton Driver improvement factor of 4, would be less than a 
4 year run. 
As mentioned before, because of the lower cross section, anti-neutrino 
running produces events in a far detector with about a factor of three 
less statistics per proton on target.  Recall that 300\,kTon-years
would correspond to 15 years (!) of running 
with the current NuMI beam configuration
and a 20\,kTon off-axis detector. 
 
Again, one would have the same $\chi^2$ behavior as for neutrino running: 
but with a significant
difference -- this time the matter effect is reversed. The reason for this is
simple: with the neutrino beam, the inverted hierarchy reduces the
$\nu_e$ appearance signal compared to the normal hierarchy and, therefore, 
in order to correctly fit the data, a larger value of $|U_{e3}|^2$ 
(compared to the one obtained with the normal hierarchy) is preferred. In the
case of the anti-neutrino beam, the inverted hierarchy enhances the
$\bar{\nu}_e$ appearance signal, and a smaller value of $|U_{e3}|^2$ is
preferred. This allows one to separate the two signs of $\Delta m^2_{23}$
if the information obtained with both beams is combined. This is what is
done in Fig.~\ref{measure_low}(bottom, left). Note that in this case
the ``wrong'' model is about sixteen units of $\chi^2$ away from the ``right''
model. It is also curious to note that, even with the wrong hypothesis,
a similar measurement of $|U_{e3}|^2$ is obtained. This coincidence, which
will not be considered too relevant, is a consequence of the fact that
the data with the neutrino and anti-neutrino beams ``pull'' the
measured $|U_{e3}|^2$ in opposite direction, and their combination meets 
somewhere ``in the middle.''

Finally, in order to determine how well the two different signs of
$\Delta m^2_{23}$ can be separated, Fig.~\ref{measure_low}(bottom, right)
depicts $\chi^2_{\rm PLUS}-\chi^2_{\rm MINUS}$ as a function of the
input value of $|U_{e3}|^2$, plus the input $\Delta m^2_{23}>0$. Note
that for $|U_{e3}|^2\gtrsim 2\times
10^{-3}$, a $\chi^2$ separation of more than 
two units can be obtained. One can turn this into an exclusion of the
``wrong'' sign by noting that, for an average experiment, $\chi^2=2$
for the ``correct'' hierarchy, 
if one combines the data obtained with the two beams (2 is the number
of degrees of freedom in this case). This implies that, for 
$|U_{e3}|^2=0.01$, the wrong hypothesis yields $\chi^2\simeq 24$, which is 
excluded at more than four sigma. A three sigma confidence level
determination of the neutrino mass hierarchy
would be obtained at $|U_{e3}|^2\simeq 0.005$, or a factor of 12 beyond 
the CHOOZ limit.  

\subsubsection{$10^{-4} < \Delta m^2_{12} < 2\times 10^{-4}$  eV$^2$}

If the best fit point to the current solar data~\cite{solar_fits} 
is close to the true solution, the
KamLAND reactor neutrino experiment will be able to not only observe 
a depletion of the reactor anti-neutrino flux, but also determine the values
of $\Delta m^2_{12}$ and $\sin^2\theta_{\odot}$ with very good precision 
\cite{Kam,Kam_sim,Kam_sim2,Kam_sim3}. 

This being the case, it is possible to determine 
$|U_{e3}|^2$ and $\delta$, and the mass hierarchy at the off-axis 
experiment. Although only one mass hierarchy is shown in the figures 
below, the effects due to the matter are significantly larger than 
those due to CP violation, and so separating the two should be 
possible \cite{minakata}.  

\begin{figure}[htp]
\centerline{\epsfxsize 13cm \epsffile{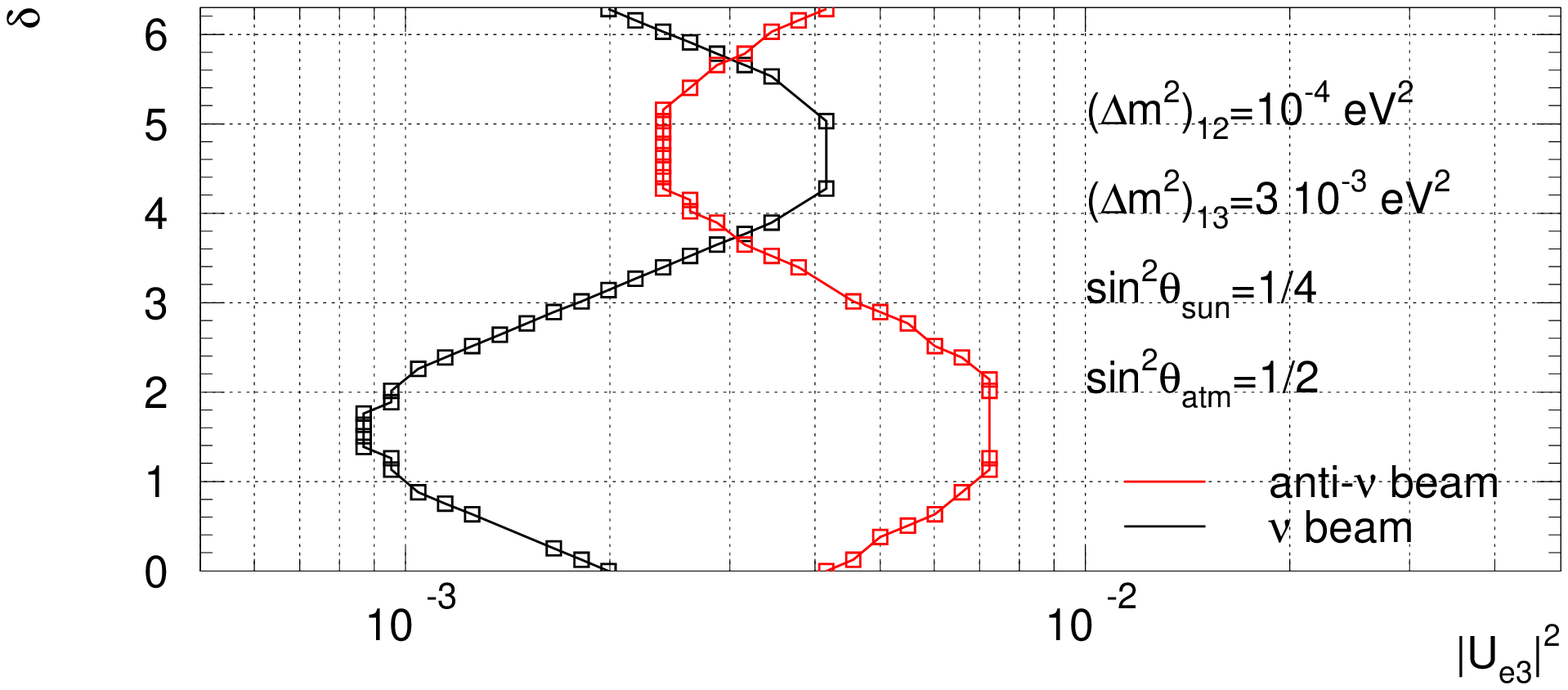}}
\caption{Three sigma sensitivity for observing a $\nu_{\mu}\rightarrow\nu_e$ signal
in the $(|U_{e3}|^2\times\delta)$-plane, 
after 120\,kTon-years of neutrino-beam running (black, darker line)
or 300\,kTon-years of anti-neutrino-beam running (red, lighter line). 
$\Delta m^2_{23}=+3\times10^{-3}$~eV$^2$,
$\sin^2\theta_{\rm atm}=1/2$, $\Delta m^2_{12}=1\times 10^{-4}$~eV$^2$, 
$\sin^2\theta_{\odot}=1/4$. }
\label{sens_lma}
\end{figure}

Fig.~\ref{sens_lma} depicts the three sigma sensitivity in the ($\delta\times
|U_{e3}|^2$)-plane for $\Delta m^2_{23}=3\times 10^{-3}$~eV$^2$ for
120 (300)~kTon-years of running with the (anti)neutrino beam. 
The sensitivity depends significantly on the CP-odd phase $\delta$, and,
as expected, the sensitivity is best for $\delta\sim\pi/2$ in the
case of running with a neutrino beam ($\delta\sim3\pi/2$ for the 
anti-neutrino beam), where the
``interference'' between the ``CP-odd term'' and the ``$|U_{e3}|^2$ term''
is constructive ({\it i.e.,}\/ one observes more events) and worse at
$\delta\sim3\pi/2$, where the ``interference'' is destructive. For smaller 
values of $\Delta m^2_{12}$, the `z-shape' and `s-shape' observed in 
Fig.~\ref{sens_lma} degenerate into vertical straight lines, such that the 
sensitivity will no longer depend on the CP-odd phase.

\begin{figure}[htp]
\centerline{\epsfxsize 14.2cm \epsffile{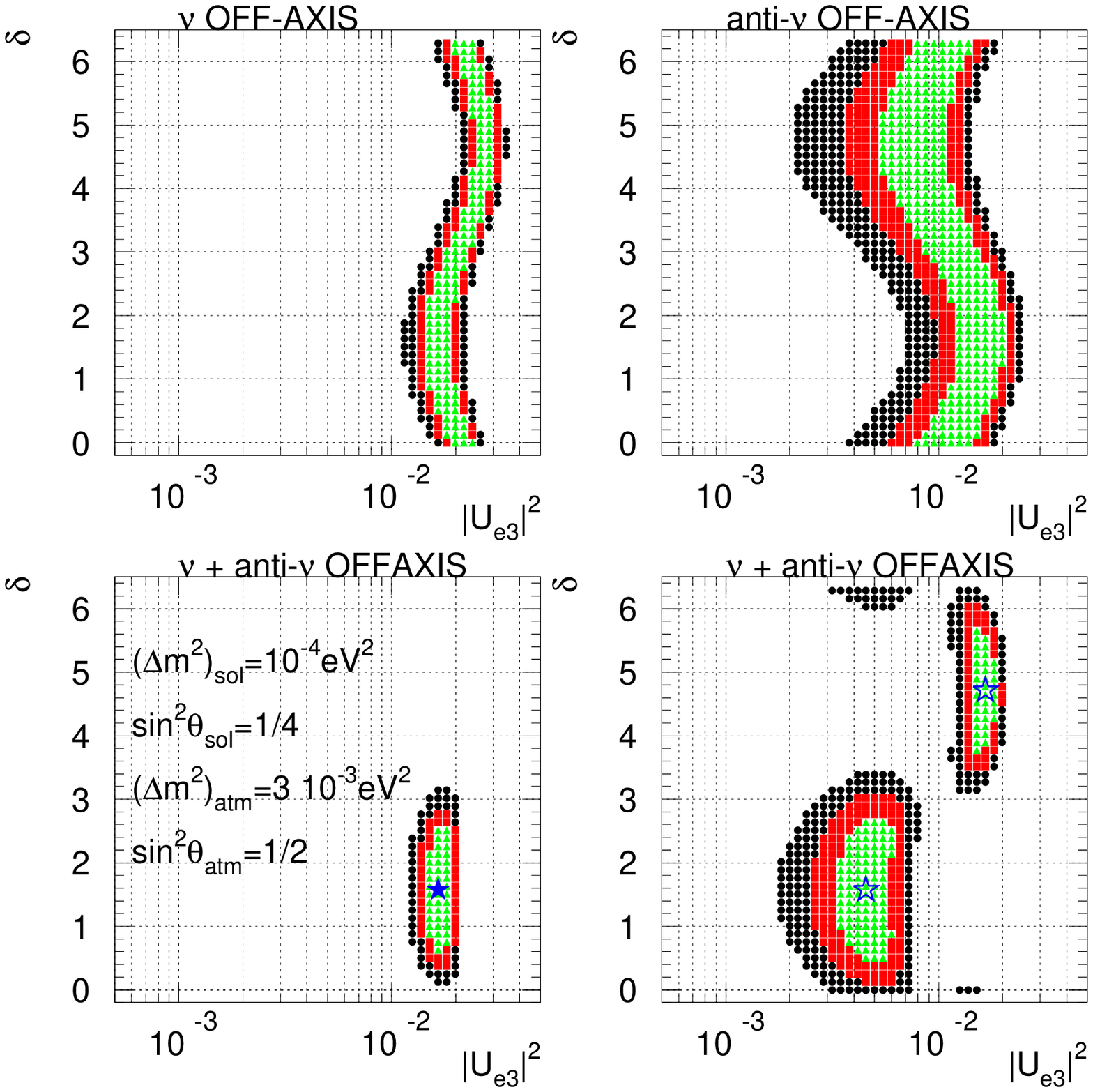}}
\caption{top -- One, two, and three sigma measurement contours in the 
$(|U_{e3}|^2\times\delta)$-plane, after 120\,kTon-years of neutrino-beam running 
(left) or 300\,kTon-years of anti-neutrino-beam running (right). The simulated data 
is consistent with $|U_{e3}|^2=0.017$ and $\delta=\pi/2$. bottom,left -- same as above,
after the two data sets are combined. The solid star indicates the simulated input. 
bottom,right -- same as before, for different simulated data points (indicated by the stars).
$\Delta m^2_{23}=+3\times10^{-3}$~eV$^2$,
$\sin^2\theta_{\rm atm}=1/2$, $\Delta m^2_{12}=1\times 10^{-4}$~eV$^2$, 
$\sin^2\theta_{\odot}=1/4$.}
\label{measure_lma}
\end{figure}

If a signal is observed, one can attempt to determine the
mixing parameters $|U_{e3}|^2$ and $\delta$.
Similar to what was done before, the atmospheric
parameters $\Delta m^2_{23}=3\times10^{-3}$~eV$^2$,
$\sin^2\theta_{\rm atm}=1/2$ will be assumed known
with infinite precision, and the same will now hold for the solar parameters
$\Delta m^2_{12}=1\times 10^{-4}$~eV$^2$, $\sin^2\theta_{\odot}=1/4$. 
Furthermore, we will also assume that
the neutrino mass hierarchy is known.\footnote{It may turn out,
for example, that table top experiments \cite{doublebeta_exp}
or the observation of supernova neutrinos~\cite{supernova}
will be able to measure the neutrino mass hierarchy.} This is done in order to 
not cloud the results presented here. Fig.~\ref{measure_lma}(top,left) 
depicts the one, two, and three sigma measurement contours in the ($|U_{e3}|^2
\times\delta$)-plane obtained after 120~kTon-years running with the neutrino 
beam.
The simulated data are consistent with $|U_{e3}|^2=0.017$ and $\delta=\pi/2$. 
One can readily note that while
$|U_{e3}|^2$ can be measured with reasonable precision, virtually nothing can
be said about $\delta$. Furthermore, the fact that $\delta$ is not known implies
that a measurement of $|U_{e3}|^2$ irrespective of $\delta$ 
is in fact less precise than what can be obtained
if the solar parameters are not in the LMA region. 

In order to improve on this picture, it is imperative to prolong our 
``experiment'' and take data with the anti-neutrino beam as well. 
Fig.~\ref{measure_lma}(top,right) depicts the one,
two, and three sigma measurement contours in the ($|U_{e3}|^2\times\delta$)-plane
obtained after 300~kTon-years running with the anti-neutrino beam (as mentioned
before, the longer running time is required in order to compensate for the 
``less efficient'' anti-neutrino beam). Again, $|U_{e3}|^2$ can be measured 
with some precision and nothing can be said about $\delta$. A comparison
of the two figures hints that a combined analyses may prove more fruitful. 
This is the case because while
the neutrino beam yields a ``z-shaped'' measurement contour, the anti-neutrino
beam yields an ``s-shaped'' contour. The reason for this is that the number of
$\nu_{\mu}\rightarrow\nu_e$ induced events 
is larger for $\delta=\pi/2$ and smaller for $\delta=3\pi/2$.
Therefore, the measurement will choose larger values of $|U_{e3}|^2$ at around
$\delta=3\pi/2$ in order to compensate for this small suppression. On the other
hand, the number of $\bar{\nu}_{\mu}\rightarrow\bar{\nu}_e$ induced 
events is smaller at
$\delta=\pi/2$ and larger at $\delta=3\pi/2$, and the opposite phenomenon
is observed.

Fig.~\ref{measure_lma}(bottom,left) depicts the result of measuring $|U_{e3}|^2$
and $\delta$ using the combined neutrino and anti-neutrino beam ``data.'' The 
situation is significantly improved, and now, a three sigma measurement of 
$\delta\neq 0$ can be performed. Fig.~\ref{measure_lma}(bottom,right) is
similar to Fig.~\ref{measure_lma}(bottom,left), except that different input
values of $|U_{e3}|^2,\delta$ are chosen. As expected, the quality of the
measurement is marginally worse for $\delta=3\pi/2$ (where $\delta$ 
is consistent
with zero at the three sigma level), and deteriorates as $|U_{e3}|^2$ decreases.
For $\delta=\pi/2$, one cannot determine that $\delta\neq 0$ or $\pi$ 
({\it i.e.,}\/ no CP-violation) at the
two sigma level if $|U_{e3}|^2\lesssim 0.004$. It should always be kept in mind that
the situation deteriorates for smaller values of $\Delta m^2_{12}$.   \\

\subsubsection{Large $\Delta m^2_{12}$} 

If $\Delta m^2_{12}\gtrsim 2\times 10^{-4}eV^2$, 
KamLAND will not be sensitive to the very rapid oscillatory pattern, and 
will only be able to observe an overall suppression of the solar neutrino 
flux \cite{Kam_sim,Kam_sim3}. In this case, the 
mixing angle $\sin^2\theta_{\odot}$ can be measured with some precision
by determining the overall suppression factor, but the value of 
$\Delta m^2_{12}$ will only be constrained to be larger than some lower limit.
An upper limit will be provided by future solar data. In order to be conservative, 
we will consider the upper bound currently provided by CHOOZ~\cite{CHOOZ}, 
and assume that $\Delta m^2_{\odot}\lesssim
7\times 10^{-4}$~eV$^2$ for large solar angles.

If this scenario turns out to be correct, precise
measurements of the atmospheric parameters and the solar mixing angle 
will probably be available, while 
$|U_{e3}|^2$, $\delta$ and the precise value of 
$\Delta m^2_{12}$ will remain unknown. Of course, a ``short-KamLAND'' or ``long-CHOOZ'' 
reactor experiment would certainly resolve this issue~\cite{miniKam}. Such
an experiment has not been proposed yet (see, however, \cite{miniKam_exp}).

What are the consequences of having a very large but poorly measured 
$\Delta m^2_{12}$? The biggest consequence, perhaps, is that even for very
small values of $|U_{e3}|^2$, a significant amount of $\nu_e$-like events
will be observed. This implies that the ``sensitivity to $|U_{e3}|^2$,''
as discussed in the two previous sections is not a particularly meaningful
quantity to study. Furthermore,
as one may fear, this will also lead to a $\Delta m^2_{12}$ versus
$|U_{e3}|^2$ ``confusion,'' (this was already alluded to in \cite{Kam_sim3}) 
similar to the one observed between $|U_{e3}|^2$
and $\delta$ in the previous section (and which continues to exist here, of 
course). In other words, a moderate $\Delta m^2_{12}$ and a large 
$|U_{e3}|^2$ will yield as many events as a large 
$\Delta m^2_{12}$ and a small $|U_{e3}|^2$. We already learned from the
previous two sections that one will be required to run both the neutrino and
the anti-neutrino beams (and accumulate enough statistics with both) in 
order to try to disentangle the three parameters.
One should keep in mind that,
while the situation is rather confusing, the number
of observed events is going to be large for very large values of the solar
mass-squared difference.

In order to address what kind of measurement one may be able to perform under
these conditions, we simulate data for $\delta=\pi/2$, $|U_{e3}|^2=0.012$,
$\Delta m^2_{12}=4\times 10^{-4}$~eV$^2$, 
$\Delta m^2_{23}=3\times 10^{-3}$~eV, 
 $\sin^2\theta_{\rm atm}=1/2$, and $\sin^2\theta_{\odot}=1/4$, 
assuming 120~kTon-years of neutrino beam running and 300~kTon-years of 
anti-neutrino beam running. As before, we assume during the data analysis that 
the atmospheric parameters, the neutrino mass hierarchy and the solar 
angle are known with infinite precision.

\begin{figure}[htp]
\centerline{\epsfxsize 13cm \epsffile{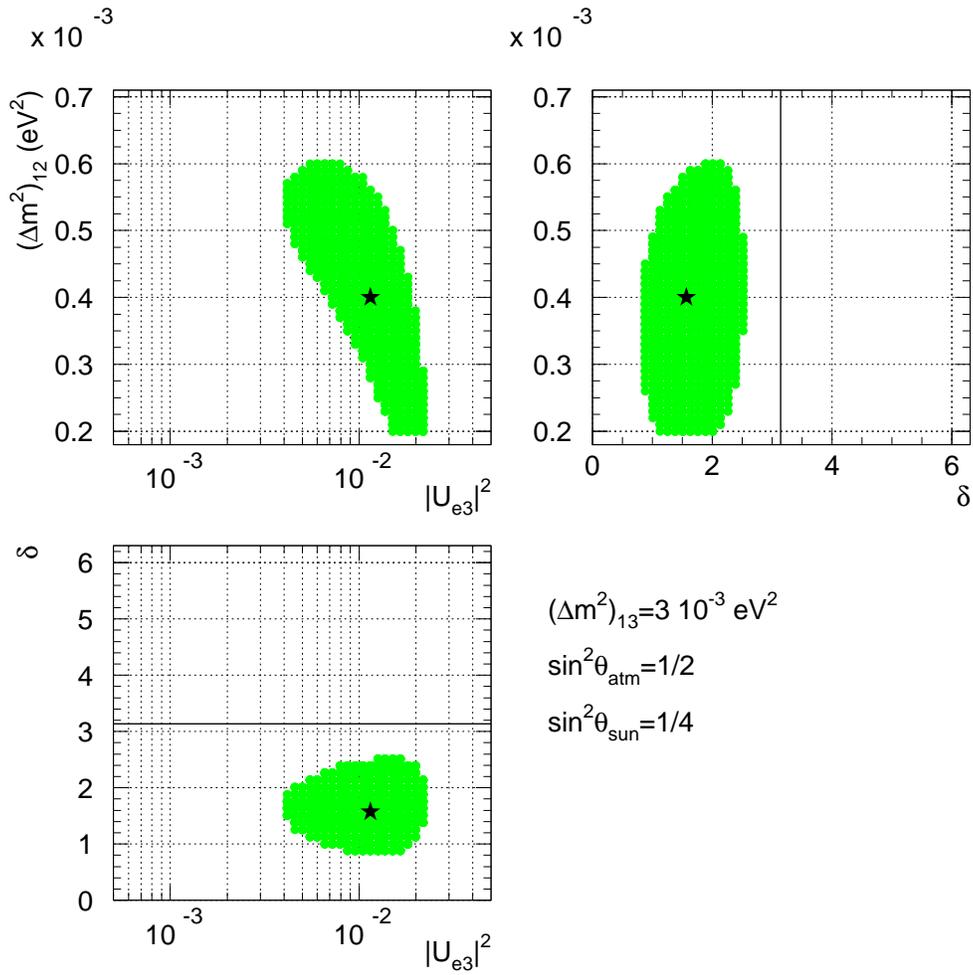}}
\caption{ Projections of the three sigma measurement surface in the 
$(|U_{e3}|^2\times\delta\times\Delta m^2_{12})$-space, after 120\,kTon-years of neutrino-beam 
running and 300\,kTon-years of anti-neutrino-beam running. The simulated data 
is consistent with $|U_{e3}|^2=0.012$, $\delta=\pi/2$ and $\Delta 
m^2_{12}=4\times 10^{-4}$~eV$^2$. 
$\Delta m^2_{23}=+3\times10^{-3}$~eV$^2$,
$\sin^2\theta_{\rm atm}=1/2$, 
$\sin^2\theta_{\odot}=1/4$.}
\label{measure_hlma}
\end{figure}

The results of the three parameter fit are presented in Fig.~\ref{measure_hlma},
where we plot the three two-dimensional projections of the three sigma 
surface in the 
($\Delta m^2_{12}\times|U_{e3}|^2\times\delta$)-space. Many comments are in
order. First of all, one should note that the solar mass difference cannot be
measured with any reasonable precision -- it is only slightly better known
than before, namely, it lies somewhere between $2\times 10^{-4}eV^2$
(the KamLAND bound) and $6\times 10^{-4}$~eV$^2$ (slightly better than 
the CHOOZ bound). The top 
left-hand panel depicts the $\Delta m^2$ versus $|U_{e3}|^2$ confusion alluded
to earlier quite well. It is curious to note however, that the capability to 
decide whether $|U_{e3}|^2$ is nonzero or not is not weak -- 
the loose constraints on $\Delta m^2_{12}$ are already 
enough to guarantee that $|U_{e3}|^2\gtrsim 4\times 10^{-3}$. Most importantly,
perhaps, at the three sigma level there is solid
evidence that $\delta\neq 0,\pi$. The reason for this is that,
for $\Delta m^2_{12}=4\times 10^{-4}$ and $\delta=\pi/2$, 
$\bar{\nu}_{\mu}\rightarrow \bar{\nu}_e$ transitions are very suppressed, 
and a zero CP-odd phase would yield far too many $\bar{\nu}_e$-like events 
when the anti-neutrino beam is on.  

\subsection{Physics Sensitivity {\em vs} Largest Mass Splitting} 
\label{Physics_Sensitivity2}
\begin{figure}
\centerline{\psfig{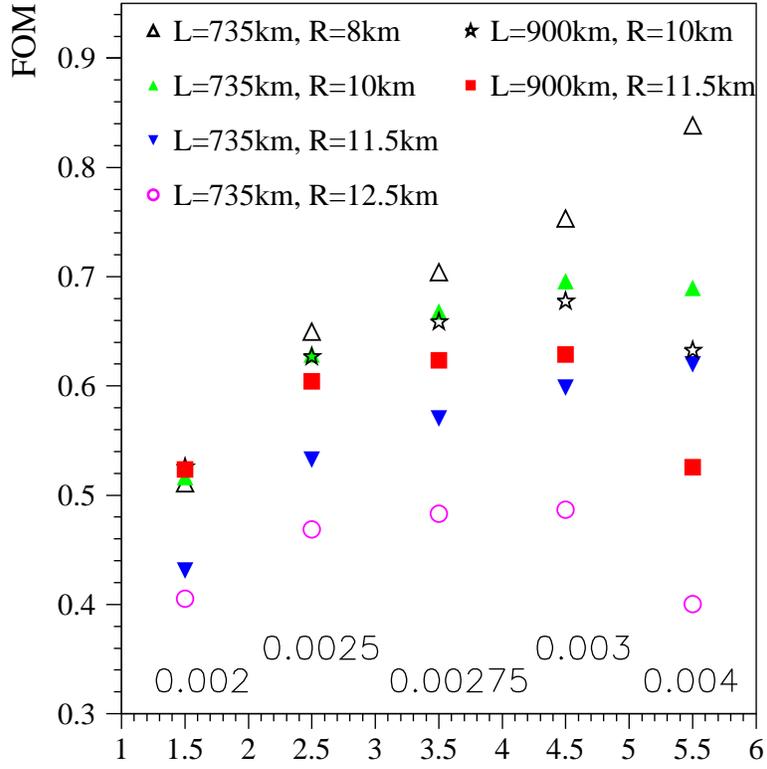}}
\caption[0]{
\label{fom}FOM for
$\nu_\mu \to \nu_e$ appearance for different values of $\Delta m^2_{13}$  
in the case of full mixing and $|U_{e3}|^2=0.01$ for different baselines 
and $\theta_\nu$, but with the same detector and
reconstruction criteria.  We only select events with visible energy between
1 and 3~GeV. If this highly segmented detector were 
at the MINOS location  
 and $\Delta m^2_{13}$ were $0.003$~eV$^2$, the corresponding FOM
would be 0.54.  
This is to be compared to the FOM for the MINOS detector that 
is only 0.39~\cite{milind}. This means, that if we ignore
the background uncertainty,  about half of the
gain in sensitivity is due to the location and the other half to the detector.}
\end{figure}

In order to conclude how much better one off-axis locations 
is with respect to each other  requires a full evaluation of
the signal and backgrounds with a realistic detector simulation and
reconstruction.  We have  performed such analysis~(Sec.~\ref{nue_aperance}),  
and we will summarize the results
by evaluating a figure-of-merit~(FOM) at each location for
$\nu_\mu \to \nu_e$ in the case of full mixing and $|U_{e3}|^2=0.01$.
We defined the FOM as $S/\sqrt{S+B}$, where
$S$ and $B$ are the signal and background events
that survive all the cuts in the reconstruction. As depicted in Fig.~\ref{fom},
off-axis experiments with angles between  $10 \le\theta_\nu\le 13$~mrad
from the axis have a  high FOM for all values of $\Delta m^2_{13}$.
The high FOM is not only due to the characteristics of the beam and
oscillation probabilities, but also due to the fact
that in all cases we can keep the NC background at the 0.5\% level,
while the reconstruction efficiency is about 40\%.
If we look in detail at the case of $\Delta m^2_{13} = 0.003~\rm eV^2$, 
we can see that the naive beam tune performed
at 735~km cannot compete with smaller $\theta_\nu$ locations at the
same baseline. This is not true for the naive beam tune performed
at 900~km, where at that location $\theta_\nu$
is already small enough to give us a high integrated flux.  We can still obtain
a 20\% increase in the FOM by reducing the baseline to 735~km
and $\theta_\nu$ to 10~mrad, but at the moment this is not particularly 
relevant, given the current lack of knowledge of neutrino oscillations 
parameters. 

\subsection{Physics Reach Summary} 

In the presence of a Proton Driver, 
our ability to obtain new physics information from $\nu_e$ and $\bar{\nu_e}$
appearance depends dramatically on the solution to the solar neutrino puzzle. It should 
be emphasized that regardless of the outcome of the KamLAND experiment, an off-axis 
experiment would provide important constraints or measurements of the lepton flavor 
sector.  

If the solution to the solar neutrino puzzle is not in the LMA region, an
off-axis experiment with the current NuMI beam line is sensitive to $|U_{e3}|^2\gtrsim 
0.003~(0.005)$ at the three sigma level for a normal (inverted) neutrino mass hierarchy. 
If one wishes to determine the neutrino mass hierarchy, it is necessary to also
take data with an anti-neutrino beam. It should be
noted that determining the neutrino mass hierarchy is not optional: the fact that the
hierarchy is unknown does not permit a precise measurement of $|U_{e3}|^2$ even if it is
sizable. 

If the solution to the solar neutrino puzzle is in the LMA region {\sl and} KamLAND is
capable of determining the solar parameters with good precision, one can attempt to 
measure both $|U_{e3}|^2$ and the CP-odd phase $\delta$. Even if the mass hierarchy is
known, measuring the CP-odd phase is not optional, as its ignorance introduces a sizable
uncertainty as one tries to extract the value of $|U_{e3}|^2$. Again we find that combining
neutrino and anti-neutrino data should provide enough information in order to discriminate
maximum CP violation ($\delta=\pm\pi/2$) from no CP violation. 

If the solution to the solar neutrino puzzle is in the LMA region but the solar
mass-squared is not well measured (as will happen with KamLAND if $\Delta m^2_{12}$
is ``too large''), one will be required to measure simultaneously $|U_{e3}|^2$, 
the CP-odd phase $\delta$, and $\Delta m^2_{12}$. Needles to say, in order to disentangle
all the different contributions to $\nu_{\mu}\rightarrow\nu_e$ transitions, neutrino and
anti-neutrino beams are required. An attempt at such a three parameter measurement is
presented in Fig.~\ref{measure_hlma}.

In order to obtain the sensitivity and 
precision discussed here, many ``kTon-years'' of data have to be collect (especially because
one is always required to run with anti-neutrinos in order to do more than just observe
and excess of $\nu_e$-like events). This can only be achieved if 1-the neutrino flux
is significantly improved with respect to the current NuMI nominal flux, 2-there is a
very massive $\nu_e$  detector on the other end with good neutral current background 
rejection capabilities.  

\subsection{Comparison with Alternative Program to Run Concurrently}
We conclude by comparing some of the results obtained here with similar
studies performed for a future JHF to Kamioka neutrino beam~\cite{JHF}.
A neutrino beam from the  JHF-proton source~\cite{JHF}  aimed at the
Super-Kamiokande detector is a neutrino project with a 295~km baseline aimed
to start in 2007-2008.
Similar to the NuMI-off-axis project, the physics goals are to measure with
an order of magnitude better precision (compared to MINOS and CNGS estimates~\cite{bar})
the atmospheric parameters ($\delta(\Delta m_{\rm atm}^2)\lesssim 10^{-4}$~eV$^2$ and
$\delta(\sin^2 2\theta_{\rm atm})\lesssim 0.01$),
confirm $\nu_\mu \leftrightarrow \nu_\tau$-oscillations or discover
sterile neutrinos by measuring the neutral current event rate, and to improve by factor of 
20 the sensitivity to $\nu_\mu \to \nu_e$-appearance.
After a five year program, the JHF-Kamioka program
should be able to exclude, at the 90\% confidence level,
$\nu_\mu \to \nu_e$ transitions for $|U_{e3}|^2>0.0015$, while at NuMI
with a 20~kTon  off-axis detector we will exclude, at the
two sigma confidence level, $|U_{e3}|^2>0.00085~(0.0015)$
with~(without) an upgrade Proton Driver (assuming $\Delta m^2_{\odot}\ll 10^{-4}$~eV$^2$
and a normal neutrino mass hierarchy). The main difference between the two programs
should come from the longer baseline proposed here, which allows the NuMI
off-axis experiment (but not the JHF-Kamioka program) to cleanly
try to address the neutrino mass hierarchy.

Underground laboratories with a  Megaton  detector are under discussion, 
but their time scale are much further away in the future.

\section{Proton Economics Issues}

  The rate at which statistics can be accumulated in many experiments scales
directly with the number of protons which can be accelerated to the
appropriate energy and delivered to that experiment. This is particularly
true of most neutrino oscillation experiments. For these experiments, one
typically builds the largest mass of detector which meets the experimental
demands and can be afforded. For world-class experiments, not only must the
detector be very massive (and expensive) but the source of protons must be
very intense. The Fermilab Main Injector currently provides a world-class
proton facility for neutrino beams. However, even the current capabilities
of the Main Injector will be stressed by the demands from the next round
of neutrino oscillation experiments, and in order to keep up with the following
round upgrades will be essential.

  Protons accelerated in the Fermilab accelerator complex serve a variety
of experimental needs. In general, the demand from multiple experiments 
requires that any given experiment may have fewer protons available for
its specific needs than would be ideally desirable. Given that neutrino
experiments can effectively use every proton which can possibly be accelerated
in order to plan the broadest physics program for Fermilab one must
also pay careful attention to the variety of demands which the future may
bring. 

In this section, we evaluate the current state of the ability of the Fermilab
complex to deliver protons, extrapolate to the ``near term'' future 
experimental program (Collider, MINOS and MiniBooNE) and make some
speculation regarding the longer-term future of the complex and proton
economics prior to the commissioning of the new Proton Driver. In some cases,
we anticipate that upgrades to the Main Injector which will also be essential
for the new Proton Driver era will already start to deliver additional proton
intensity even before the Proton Driver itself will be commissioned. We
believe that this presents a highly attractive and sensible investment path
for the laboratory.

\subsection {The Current and Near-Term State of the Economy}

Currently, the Main Injector is running primarily
for anti-proton production for the collider. In the current mode of operation,
a single batch of $\approx 4.5\times 10^{12}$ protons is first accelerated
in the Booster to 8~GeV and then injected into the Main Injector and
accelerated to 120~GeV before being delivered to the anti-proton production
target. The current cycle for the Main Injector is 2.3 seconds. The resulting
number of protons accelerated per year is about $2\times 10^{19}$, in both
the Booster and Main Injector. 

Starting immediately, the demand for protons
accelerated in the Booster will go up dramatically with the commissioning
of MiniBooNE. For MiniBooNE, six Booster acceleration cycles will be
executed for every one Main Injector cycle and the resulting protons will
be extracted to the MiniBooNE target. It is worth noting that the Booster
would nominally be capable of delivering almost twice that number of protons
except it will currently be limited due to irradiation of critical devices
due to proton losses in the machine. The MiniBooNE target could certainly 
handle the additional protons and additional protons would be very beneficial
to the experiment. 

In 2005, MINOS will also begin to run and place yet more
demands on both the Booster and Main Injector. MINOS requires that the 
Main Injector run in ``Multi-Batch'' mode where six bunches of protons are
injected from the Booster in every MI cycle. At the same time, because of 
physics demand for MiniBooNE to operate in anti-neutrino mode as well as
neutrino mode we anticipate that those experiments will run simultaneously
placing additional demands on the Booster. We note that the nominal plan
for MINOS running calls for $3.8\times10^{20}$ protons per year 
delivered to the NuMI target, roughly 15 times the number of protons 
currently accelerated to 120~GeV. At the same time, the protons delivered 
for anti-proton production is also roughly planned to double from the 
current number for Run IIB.  The number of protons which will have to
be accelerated through the Booster will have to be 30 times what is 
currently routinely accelerated. One final consideration is that the laboratory
is currently building the ability to extract Main Injector beams for
test beams and experiments (one of which is already approved, E907) in the
Meson area. These experiments will typically not increase the demand for
protons, but affect the overall economics by requiring an extended 
Main Injector cycle for slow extraction.

The first step in moving towards the future has already been taken in 
preparations for the initial running of MiniBooNE. Upgrades to the
shielding around the Booster have been added at some locations in order to
ensure that external radiation limits will not reduce the number of protons
which can be accelerated. In addition, beam ``notching'' has been introduced
to reduce the exposure of critical devices to radiation upon extraction.
Even with these improvements, the number of protons which can be delivered
to MiniBooNE over the next year or so will be limited not by any intrinsic
features of the Booster, but rather by proton losses causing the machine to
become too radioactive. Work is underway to improve that situation by adding
additional RF controls and strategic collimation where protons will be lost
rather than in critical devices such as RF cavities. In order for MiniBooNE
and MINOS to run simultaneously yet another factor of two improvement will be
needed in the loss of protons compared to anything which is currently planned
for MiniBooNE running.

Once MINOS begins running, the Main Injector must run in ``multi-batch'' mode.
Currently, multi-batch mode is in a distinctly developmental status. 
Multi-batch operation of the Main Injector was briefly demonstrated when it
was first commissioned in 1999. For that demonstration, six batches
containing a total of $2 \times 10^{13}$ protons were accelerated to 120~GeV.
For a variety of technical reasons, it is not currently possible to replicate
this operation, but with some investment in the complex it is anticipated
that this should certainly be feasible again within the next year or so.
In December of 2001, the study of  multi-batch acceleration was once again
undertaken by a group interested in studying and developing the capabilities 
for MINOS. The current ability to accelerate protons in multi-batch mode is
limited to about $1.8 \times 10^{13}$ protons per cycle. Main Injector experts
believe that improvements in the RF feedback and damping will permit
multi-batch operation to accelerate up to $3 \times 10^{13}$ protons per
batch by 2005 with one (of six) of those batches going to anti-proton production
and the remaining five going to NuMI/MINOS. Note that achieving even this
intensity of protons requires that the average number of protons in a Booster
batch must also be $5 \times 10^{12}$. 

It should be noted that at present there is no defined plan that will 
manage to deliver all of the protons that will nominally be demanded starting
in 2005. In order to even reach the ``current goals'' set by Beams Division
a significant amount of work is required and to meet the nominal demands
from the combination of Collider, MINOS, MiniBooNE and a modest test-beam
program will require roughly a factor of two additional improvement beyond
that. Although not yet defined, a study group has been commissioned to evaluate
possible incremental improvements in the current accelerator complex which
may be able to deliver this additional factor and beyond over a period of 
several years. In the following section, we anticipate some of the conclusions
which that group will offer.

\subsection {The MINOS+Collider Era, 2005 to 2008}

Starting in 2005, the Fermilab accelerator complex will face an unprecedented
demand for numbers of protons. In just one year, the demand for protons 
from the Booster and Main Injector\,(MI)
will equal the number of protons delivered to fixed target experiments
in the entire first 30 years of
the laboratory. Current beams division planning assumes the following
operating scenario for this era:
\begin {itemize}
\item Booster operation will be improved so that a total of $5\times 10^{12}$
protons will be delivered per batch.
\item Main Injector operation will be improved so that a total of six
batches (each with $5\times 10^{12}$ protons) can be reliably accelerated
with a cycle time of 1.9 seconds.
\item A total up-time of $1.8\times10^7$ seconds of production acceleration
cycle per year will be realized for a total of $\approx 2.8 \times 10^{20}$
protons accelerated to 120~GeV per year. 
\item One-sixth of those protons will go to anti-proton production and
five-sixths will go to the NuMI target. No significant running is currently
planned (by Beams Division) for MiniBooNE running and it is assumed that
running for other fixed target experiments and/or test beam will impact
the total protons accelerated by no more than 10\%. (This may be come in
a number of different forms.)
\end {itemize}

Although the above scenario appears realistic, we note that the program is
already going to be short of nominal expectations and anticipate that the 
following conflicts are likely to arise:
\begin {enumerate}
\item Under the above scenario, the number of protons delivered to the NuMI
target will be only $2.5 \times 10^{20}$ per year rather than the nominal
design plan of $3.8 \times 10^{20}$ per year.
\item A conceptual plan exists for slip-stacking one batch of protons into
the Main Injector to increase the anti-proton production. The slip-stacking
will increase the cycle time of the Main Injector one implemented and will
cause an additional 10\% reduction in the protons for NuMI.
\item No running is planned for MiniBooNE. Either a plan to accelerate more
protons in the Booster by a factor of two must be developed or running MINOS
and MiniBooNE simultaneously will reduce the proton intensity to the NUMI
beamline 
by a factor of two. We anticipate that extended running of MiniBooNE is likely
to be requested and recommended by the PAC.
\item Should a substantial test-beam program be undertaken one may anticipate
an additional 10\% reduction in protons delivered to NuMI.
\end {enumerate}

Hence, we observe that in order to meet nominal demands for protons in this
era, a factor of two improvement will be needed in the number of protons
which can be accelerated to 8~GeV in the Booster compared to any current
plan and that a factor of two improvement will also be required in the
ability to accelerate protons in the Main Injector to 120~GeV from any 
existing plan. Furthermore, in this time-scale it is clear that these
improvements will not come from commissioning of the new Proton Driver.

Fermilab and the MINOS collaboration are interested in possible
modest investments which can be made in the existing accelerator complex to
increase the proton intensity in this timescale. It will be particularly
attractive to invest in hardware which will also be necessary in the 
longer term should a new Proton Driver be built. Appendix \ref{appendA} 
lists possible improvements that could be made by the time NUMI 
starts receiving protons on their target.  

The bottom line is that it appears that increasing the proton intensity
within the existing accelerator complex will certainly be possible within
the timescale of 2005-2008. Roughly, one might categorize the expected
improvements based on level of investment and effort for a particular
proton intensity. Table \ref {tab:protonintensity} summarizes the
approximate situation that can be expected based on a particular level of
investment.

\begin {table} [tb]
\begin{center}
\begin {tabular}{|l|l|l|l|}
\hline
Investment level  &  Very Rough Cost & 120~GeV Protons  & 120~GeV Protons \\
                  &                  & in 2005          & in 2008         \\
\hline
$\approx$None     &  $\approx$ \$0 & $1.3\times10^{20}$ & $1.3\times10^{20}$ \\
Small      &  $\approx$ \$2-4  M  & $2.7\times10^{20}$ & $3.0\times10^{20}$\\ 
Medium     &  $\approx$ \$10  M  & $3.0\times10^{20}$ & $4.5\times10^{20}$\\
Substantial & $\approx$ \$50M     & $3.5\times10^{20}$ & $9.0\times10^{20}$\\
\hline
Ultimate & $\approx$ \$200M     & --- & $16.0\times10^{20}$\\
\hline
\end {tabular}
\end{center}
\caption { \label {tab:protonintensity} The total protons per year which
can be expected to be accelerated to 120~GeV for several different levels
of investment in the existing accelerator complex. It will take time for
some of the improvement programs to be carried out. Numbers are shown for
2005 and 2008 assuming an adiabatic investment. Note that these are the
total protons which are accelerated, some of which go to anti-proton 
production, some to NuMI and a few for other purposes. Note that the
investment levels all include an assumption of twice the current number 
of protons accelerated to 8~GeV in the Booster than currently needed just
for MiniBooNE operation. MINOS expectation is $4\times10^{20}$ protons
per year. The last line gives the expectations for the most economical
Proton Driver design that is available at the moment~\cite{booster,pdriver}.} 
\end {table}

\subsection {2008 and Beyond; Towards the Superbeam}

   The physics program of Fermilab becomes less well-defined in this
era. We note that the following experiments may all exist at this time
with some demands for protons and/or demands on modification of the
acceleration cycle from the fastest possible:
\begin {itemize}
\item BooNE (or other follow-on Booster neutrino experiments): Needs only
Booster or new Proton Driver protons.
\item CKM: Requires slow-spill extraction from the Main Injector. This has
a significant impact on cycle time for experiments which do not require
a slow spill.  
We assume that a ``reasonable compromise'' will be that the cycle
average cycle-time reduction for anti-proton production and neutrino
experiments might be about 30\%.
\item BTeV: Requires anti-proton production as for the current/future
collider program.
\item MINOS: We anticipate that regardless of a new neutrino experiment 
that it will be interesting to extend running of MINOS either with
anti-neutrinos or for extended precision with neutrinos.
\item New Off-Axis NuMI Experiment: As discussed elsewhere in this 
document we anticipate this to be a new important part of the Fermilab
program. But it makes no additional demands on protons from MINOS!
\item Test beams and other uses: We assume that these uses will be limited
to no more than 10\% impact on the high-rate proton production. Most likely,
it can be parasitic to the impact of CKM and hence we do not include it
in the impact on high-rate proton production.
\end {itemize}

Hence, the demand for protons goes up in this timescale, making relatively
fewer protons available for each experiment. Using the ``medium investment''
estimate for proton acceleration from Table~\ref{tab:protonintensity}
we find that protons available for neutrino experiments in this era will
remain below what the nominal proposal expectations set for the MINOS
experiment. Specifically, the MINOS expectation is $4\times10^{20}$ protons
per year, corresponding to Main Injector power of 0.4 MW. However, the
number of protons which will be delivered to the NuMI target in 2008
if all of the above conditions exist will be just $2.6\times10^{20}$.
Given that the demands for anti-neutrino running for MINOS and/or better
reach in an Off-Axis detector call for yet more protons to the NuMI target,
it is clear that some additional investment in proton acceleration is
essential if all of these programs are to run simultaneously. We see that
significant investment in the existing complex can in fact deliver more
protons, but the demand from the remaining program still means that 
only $5.2\times10^{20}$ protons per year can be delivered to the NuMI target.
We note that this number is based on a rather large new investment in
the existing accelerator complex, some of which will be attractive anyway
if one anticipates a new Proton Driver. Although even more dramatic 
investments in the ``existing'' complex are possible (superbunch 
acceleration in both the Booster and Main Injector for instance) these 
start to become rather expensive and begin to be on the class of 
a new Proton Driver in order to go very much beyond the intensities that
the ``significant investment'' level affords.

The ultimate beam for a Superbeam program at FNAL will require a new
Proton Driver.
The Proton Driver design described in~\cite{booster,pdriver} will allow us
to bring the NuMI neutrino beam power up from 0.4~MW to 1.6~MW.
This design  is based on an 8~GeV circular machine with a circumference
of 473.2~m, and it will provide $2\times 10^{13}$ protons per
pulse instead of the assumed $5\times 10^{12}$
of the current booster.
In addition, the total luminosity could be further increased by 30\% if
the current linac gets a 200~MeV upgrade. In this case,  we would get
$3\times 10^{13}$  protons per pulse.  This machine is estimated to cost
US\$160M.

An  alternative design made out of only a  linac to accelerate
protons up to  8~GeV,
using SNS and Tesla style superconducting cavities, is  also under
investigation~\cite{foster}. 

In either case, changes to NuMI will be required, and are listed in 
appendix \ref{appendB}, but the possibility
of having a broad physics program that includes projects  like NuMI
and Boone running concurrently becomes realistic.

\section{Conclusion}

With the turn-on of the LHC in the next decade, Fermilab will relinquish its
long-held position as the energy frontier in particle physics.  However,
given the considerable expertise that will still be resident here, and the
experimental infrastructure that is currently being built, Fermilab can 
make a transition into being the intensity frontier of particle physics, 
and can take advantage of
the considerable investment this field has already made in this laboratory.
By upgrading the proton source that currently supplies both the collider and
neutrino programs, Fermilab can not only take the next important step in 
neutrino
flavor physics measurements, but can also extend its reach in anti-proton
studies, and possibly the the host of a new cold neutron source.

While currently the most outstanding issue of neutrino flavor physics is the
size of the last unmeasured mixing angle, $\theta_{13}$, seeing evidence for
it is only the beginning of the story.
The ultimate goals for neutrino flavor experiments
are to determine the neutrino mass hierarchy, and if possible measure
leptonic CP violation (assuming the solution to the solar neutrino
puzzle lies in the LMA region).  Both generations of experiments require a
very intense neutrino beam to be able to use a long baseline,
preferably with a very narrow energy spectrum to reduce the considerable
backgrounds associated with these measurements.  The NUMI facility which is
currently being built can provide such a beam, but a new detector sensitive to
$\nu_e$ appearance is required.  To get to the second generation of experiments,
namely the determination of the mass hierarchy and search for CP violation,
a very intense anti-neutrino beam is required, and in order to make up for the 
lower
anti-neutrino cross section.  An upgrade of the proton source at Fermilab
could be the step which finally enables us to know the number of
heavy neutrinos, and ultimately whether or not
there is CP violation in the lepton sector.

\clearpage 
\appendix

\section{Possible Strategies for Improving Proton Source in Near Term} 
\label{appendA}

The following is an
annotated list of issues currently under discussion for possible improvements
in the accelerator complex on this time scale:
\begin {enumerate}
\item LINAC: Upgrades to the existing LINAC offer no direct increase in
intensity in the complex but some improvements could offer indirect benefits.
  \begin {itemize}
  \item Energy increase to lower space-charge effects at injection
      to Booster? This could reduce proton losses and permit an extracted
      beam with smaller phase-space. The end result could be an increase
      in the total number of protons which can be accelerated in the Booster
      as well as better-behaved beams for stacking purposes. The cost
      is relatively high.
  \item Chopped beam for cleaner capture in Booster: This could reduce proton
      losses, allowing more acceleration cycles per year. This is not a very
      developed concept.
  \end {itemize}
\item Booster: Proton acceleration in the Booster is currently limited by
unstable beam at high intensities leading to radiation from proton losses
and extracted beam with poor properties for additional acceleration. Booster
upgrades can both directly increase intensity and make for higher MI 
intensity by providing clean beams for stacking and easy acceleration.
Limitations in the direct intensity increase are not completely understood
but many experts estimate that it should be possible to increase intensity
from the current typical operating mode, $\approx 4.5\times10^{12}$ protons
per batch, to $\approx 6-7\times10^{12}$ protons per batch. The cost of
most of these systems for the Booster are very modest compared to the
cost of a new Proton Driver, but the returns are correspondingly modest.
   \begin {itemize}
   \item Reduction in space-charge effects at injection by making longer
         bunches or adding electron shielding? 
   \item New hardware to help stabilize the beam and reduce proton
         losses including:
         \begin {enumerate}
         \item Ramped correctors
         \item New RF damping hardware
         \item New collimators
         \item Larger RF cavities
         \item Inductive inserts
         \item Additional acceleration RF harmonic cavities
         \end{enumerate}
   \end {itemize}
\item Main Injector: Improvements in the Main Injector fall into three
      main categories; decreasing the cycle time, improvements to permit
      the machine to handle more total proton flux, regardless of source,
      and proton stacking schemes. As the total current of protons in the
      machine is increased, it will be necessary to add extra RF power and
      damping under any circumstance, even with a new proton source. Hence,
      much of that investment can be viewed as ``on the path'' of the
      proton source. 
   \begin {itemize}
   \item Additional RF power to handle extra proton intensity and permit
         reduction in cycle time. Reduction in cycle time could be as much as
         25\% with no modification to existing RF cavities.
   \item Reduction in cycle time by ``tuning'' the acceleration cycle. This
         probably would require very little new hardware and could result in
         a reduction in cycle time of 10-15\%.
   \item New RF damper electronics and components. This would be 
         necessary to go to higher
         intensity and more sophisticated and expensive as the intensity goes
         even higher.
   \item Inductive inserts to improve stability.
   \item Collimators to protect critical components from beam losses.
   \item Yet more RF power with cavity modifications and/or new cavities 
         to reduce the ramp time even more and the total cycle time
         by as much as a factor of 2.
   \item Slip stacking: Unlikely to be of interest for multi-batch operation.
   \item Barrier RF stacking: Appears promising for increasing protons
         accelerated to 120~GeV by 60-70\%. Requires well-behaved Booster
         beam and new barrier RF systems in Main Injector. Operates on
         principles already in use in the Recycler.
   \end {itemize}
\end {enumerate}

\clearpage

\section{NuMI Changes Required in Case of PD Upgrade}

\label{appendB} 
The NuMI Beamline is designed to handle ``only'' 0.4\,MW of proton
power, and will itself require modifications to allow it to accept
five times the integrated proton power.  A detailed description of
the modifications required can be found in the Proton Driver machine
study~\cite{Fermilab-TM-2169}.
After a description of the beamline, we give here only a brief
synopsis of those modifications.  In fact, the modifications required
depend very significantly on what the ultimate increase in proton
power is.  If one were to run NuMI with twice the proton
power, the modifications would not need to be as extensive.
Furthermore, several of the modifications can be
avoided by changing the proton beam spot size, which although not 
trivial, is feasible.  

The neutrino beam at NuMI is created when 120\,GeV protons from the Main
Injector strike a 0.94\,m graphite target located roughly 40\,m below ground
in the NuMI tunnel.  The secondary pions and kaons are then focused
by two horns towards a 670\,m decay pipe.  The uninteracted protons and
particles that did not decay hit the hadron absorber, located about 725\,m
from the target.  Finally, beamline monitors
measure the distribution of particles both upstream and downstream of the
hadron absorber, and also in alcoves embedded in dolomite farther downstream.
The target, the horns, and the decay pipe itself all are water cooled, and
even for 0.4\,MW the entire beamline requires an impressive amount of
shielding due to radiation concerns.

The modifications necessary for the primary beam optics are the most
trivial, assuming a cycle time at or near 1.87\,seconds.
Because the beamline is designed to match the dynamic aperture
of the Main Injector, whatever the Main Injector
supplies, the beamline can take.
However, the losses per minute must be maintained at the same level as
those with 0.4\,MW proton power, so additional collimation may be required.

The modifications necessary for the NuMI target and horns depend critically on
the spot size of the incoming proton beam.  With a factor of 5 more
intensity the NuMI target would be too hot after just a few pulses, with
the current proton spot size.  However, if both the proton beam spot size
were increased by a factor of three above the current design, and the target
itself was a rod three times the beam width, then the same material could
withstand the increase in proton power, with similar cooling design.
In fact, the current NuMI target is taller than it is wide, so it is
estimated that if one had just twice the proton spot size in the vertical
direction, and twice the proton power
the target would not need to be upgraded.
Increasing the proton spot size also means the decay pipe windows would not
have to be modified from their current designs.
The NuMI horns themselves suffer the most from the repeated pulsing, not
from the passage of particles through them.  Thus they could withstand
an increase in power.
However an increase in the repetition rate would change the
expected horn's lifetime.  The current lifetime expectancy is 1\,year, for a
1.87\,second repetition rate.

The modifications for cooling the beamline are the most severe.  The
target area is currently cooled by a very high air flow rate.
If there were 5 times the proton power the region would
have to be water cooled instead.
The decay pipe itself is currently being cooled by water flowing through
pipes that run along its entire 670\,m length.  The section that would require
additional cooling is less than a third of this, however.  One could
conceivably cut holes in the concrete shielding and add cooling pipes
in that area.  Finally, the hadron absorber
itself would also need to be modified to allow more cooling, although
the modification is not severe--the core is currently water cooled, only
the first block of steel which is not cooled now would need to change.
All of these upgrades require modifications to what will be highly radioactive
components.

Finally, the increase in proton power will also require a new analysis of
radiation safety issues.  The issue of proton beam losses remaining low
was addressed earlier, and if those proton beam losses are maintained,
other radiation safety issues will be quite manageable.
These issues are currently being addressed
using extrapolations from other data, which necessarily bring in
significant uncertainties, which translate into large safety factors.
It is expected that the measurements of the water pumped from the NuMI
tunnel and released to the surface waters will be greater than a factor
of ten below the surface water limits.  Similarly it is expected that
measurements from NuMI's monitoring wells will be negligible.  Extrapolating
from these measurements would show an upgrade in a factor of 4 in intensity
to be feasible.  As for airborne activation,
it is likely that the target shielding  and hadron absorber will need additional
air seals, but given the new cooling that will also be required, this
would be a minor additional cost.

In summary, the most significant upgrades required in the NuMI beamline
would be those associated with cooling and sealing the target hall,
and a new target with cooling.  The decay pipe and absorber cooling
would be the next most costly changes.  Radiation safety issues do not
drive any of the costs.  The overall cost for a NUMI upgrade to accept 
4 times the
intensity and a 1.5 second repetition rate would range between 4 and 16
million dollars, which, although substantial, is less than 10 \% of the 
cost of a new proton source.

\end{document}